\documentclass[aps,prb,twocolumn,superscriptaddress]{revtex4-2}
\usepackage{amsfonts,amssymb,amsmath,bm}
\usepackage{graphicx,graphics,float}
\usepackage[dvipsnames,svgnames,x11names]{xcolor}
\usepackage{xr-hyper}
\usepackage{hyperref}
\usepackage{tabularx}
\usepackage{microtype}

\hypersetup{
    colorlinks=true,
    citecolor=Fuchsia,
    urlcolor=Blue4,
    linkcolor=TealBlue,
}

\setcitestyle{super}
\makeatletter

\renewcommand\frontmatter@abstractwidth{\dimexpr\textwidth\relax}
\makeatother

\makeatletter
\let\old@bibcite\bibcite
\renewcommand{\bibcite}[2]{}
\makeatother
\makeatletter
\let\bibcite\old@bibcite
\makeatother

\newcommand{\supp}{\textbf{Supplementary Information}}

\begin{document}

\title{Scalable Simulation of Quantum Dynamics on Topological Quantum Hardware}
\author{Kritanjan Polley}
\email{kritanjan251@gmail.com}
\affiliation{Simons Center for Theoretical Computational Chemistry, New York University, New York, New York 10003, USA}
\author{Mark E. Tuckerman}
\email{mark.tuckerman@nyu.edu}
\affiliation{Department of Chemistry, New York University, New York, New York 10003, USA}
\affiliation{Simons Center for Theoretical Computational Chemistry, New York University, New York, New York 10003, USA}
\affiliation{Department of Physics, New York University, New York, New York 10003, USA}
\affiliation{Courant Institute of Mathematical Sciences, New York University, New York, New York 10012, USA}
\affiliation{NYU-ECNU Center for Computational Chemistry at NYU Shanghai, Shanghai 200062, China}
\date{\today}

\begin{abstract}
Quantum computers offer a significant advantage in simulating quantum systems compared to classical computers for certain problems, although most current applications are limited to calculating static molecular properties using hybrid quantum-classical hardware. In this work, we establish a framework for the representation of quantum dynamics in molecular and condensed matter systems, designed for execution on topological quantum hardware. By leveraging the non-Abelian braiding statistics of Fibonacci and Ising anyons, we utilize the Solovay-Kitaev algorithm to  approximate unitary propagators for a range of systems. We demonstrate the efficacy of these algorithms across a hierarchy of complexity, from two-level systems and one dimensional double-well potentials to condensed phase spin-boson models, simple molecules, and molecular reaction kinetics. These algorithms provide a scalable and robust pathway for simulating many-body condensed phase chemical physics on fault-tolerant quantum devices.
\end{abstract}

\maketitle

\section{Introduction:  Anyons \& Topological Quantum Computers}
Quantum computers offer polynomial or even exponential improvements over the best known classical approaches for certain problems in quantum physics.~\cite{Feynman82,Mermin_2007,Aaronson08,martonosi19} Their primary advantage lies in their ability to coherently manipulate an exponentially large Hilbert space. Simulating the real-time dynamics of many-body quantum systems is a cornerstone of contemporary chemical physics, yet it remains computationally prohibitive for classical hardware. Quantum gates are designed to implement unitary propagators that evolve quantum states in real time efficiently, which is a particularly interesting computation for generating spectral functions, response functions, reaction rates, scattering, transport, and strongly correlated dynamics.~\cite{daskin11,tacchino20}

Decoherence of spin states and their sensitivity to errors is a significant challenge in quantum two-state, spin based, or bosonic quantum computers, although important progress is being made in error-correction approaches.~\cite{ECC} Topological quantum computation offers a promising framework for constructing fault-tolerant quantum computers.~\cite{nayak08,freedman03,kitaev03,das06,stern13} It relies on topological states of matter with quasiparticle excitations in two spatial dimensions referred to as ``anyons" due to the continuum of phase values that their wave functions can acquire upon particle exchange, $\psi(\bm{r}_2,\bm{r}_1)=e^{2i\theta} \psi(\bm{r}_1,\bm{r}_2)$, ($\theta\in [0,\pi]$).~\cite{xue26} A particular class of anyons, namely the non-Abelian anyons, obey non-Abelian fusion rules and braiding statistics~\cite{nayak08}. The Hilbert space used for topological quantum computation is the subspace of the full Hilbert space of the system and is comprised of the degenerate ground states with a fixed number of quasiparticles at fixed positions.~\cite{StanescuBook} The associated topological gate operations on this Hilbert space are performed by braiding the word lines of the associated anyons and require only that the topology of a braid sequence be maintained via a specific order of braids, which is largely protected against local noise and, therefore, inherently fault tolerant.

This exotic topological state of matter arises in low dimensional (2D) materials, at the interface between a semiconducting material and a conventional superconductor at very low temperatures.~\cite{haldane17} On the surfaces of such 2D materials, quasiparticles called Majorana zero modes are created.~\cite{sarma15,lutchyn18} Anyons exhibit specific evolutionary scenarios; for instance, they can be created or annihilated in a pairwise fashion, they can be exchanged adiabatically, they can be fused to form other types of anyons (a process known as fusion). Non-Abelian anyons are characterized by the existence of multiple fusion outcomes. The initial state preparation of the Hilbert space is carried out through fusion operations. When two anyons are exchanged, or braided, the state in the fusion space undergoes a unitary evolution.~\cite{xu24braid}

Initializing a topological quantum computer requires the definition of an $n$-qubit computational space, achieved by generating anyons from the vacuum and stabilizing their positions. This process establishes a protected, non-local fusion space that serves as the system's computational manifold. The computation concludes with a readout phase where, because computational basis states correspond to distinct pairwise fusion patterns, projective measurements are executed by physically fusing anyon pairs.~\cite{Bonderson08} The resulting outcomes, dictated by the specific anyon classes, ultimately reveal the final state of the system.

\begin{figure}[!b]
    \centering
    \includegraphics[width=\linewidth]{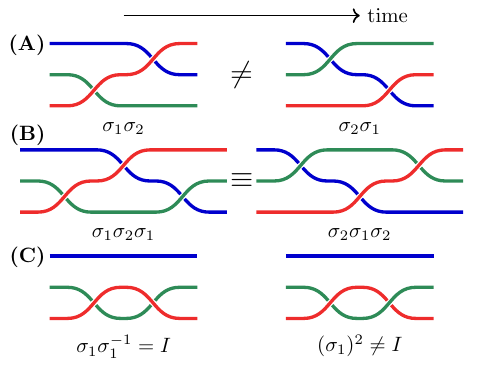}
    \caption{Diagrammatic representation of the properties of the braid generators and braid operations. (A) shows the non-Abelian nature of the braid groups. (B) shows the Yang-Baxter relation (\textit{c.f.} Eq.~\eqref{YangBaxter2}). (C) displays the counterclockwise rotation for the braid generator $\sigma_i$'s that makes $\sigma_i^2\neq \bm{I}$.}
    \label{figBraidDiagrams}
\end{figure}

\begin{figure*}
    \centering
    \includegraphics[width=\textwidth]{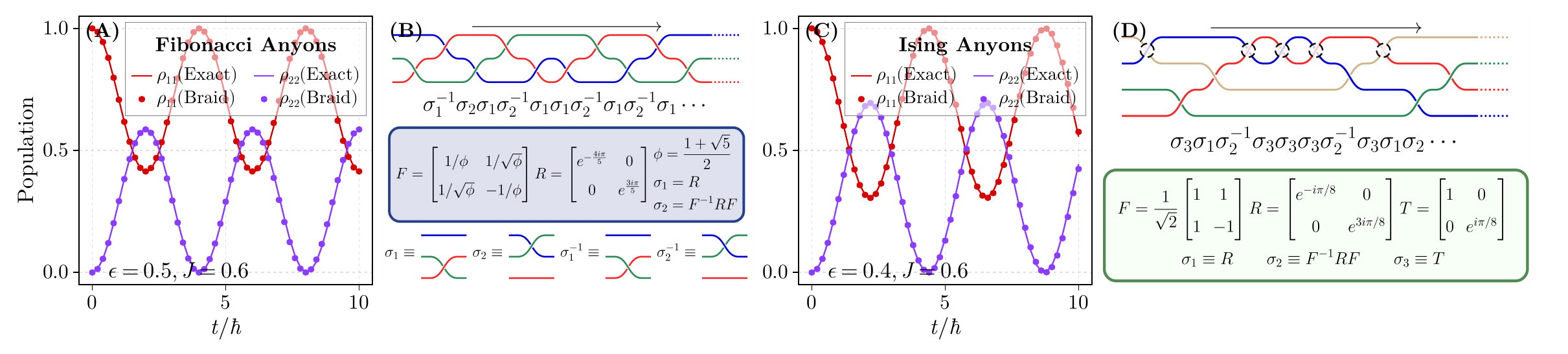}
    \caption{Braiding sequences generated for a two-level system using the Solovay-Kitaev algorithm are employed to simulate the time propagation of the system. In panel (A), the exact propagation of population dynamics, obtained by directly exponentiating the Hamiltonian, is depicted with solid lines, while the result obtained from Fibonacci anyon braiding is shown with scattered dots. The first 10 elements of the Fibonacci braiding sequence are displayed above panel (B). Panel (B) also shows the matrix elements for the braiding operations and the individual braiding diagrams for 4 gates; $\sigma_1$, $\sigma_2$, and their inverses. Panel (C) presents the population dynamics of a similar Hamiltonian, where we utilized Ising anyon braiding. The first few braiding sequences and the matrix elements of the braiding elements are shown in panel (D). The set of gates for Ising anyons possess a non-topological gate, specifically a rotation gate ($T$ in the figure), which is represented by a dashed circle in the braiding diagram to stress that the rotation is not a topological gate. $\Delta t=0.1\hbar$ is used in both cases, the braiding word above approximates the unitary propagator $e^{-i\Delta t H/\hbar}$ with infidelity $10^{-7}$.}
    \label{fig2LS}
\end{figure*}

Fibonacci anyons~\cite{minev25,trebst08} are the simplest non-Abelian anyons with only one anyon particle, denoted $\tau$, that satisfies the fusion rule $\tau \times \tau = 1 \oplus \tau$, where $\oplus$ denotes two fusion channels, and ``1" denotes the vacuum. They are also a model for universal quantum computation. On the other hand, Ising anyons, denoted $\sigma$, have two fusion channels, denoted by $\sigma \times \sigma = 1 \oplus \psi$, where $\psi$ is a fermion that is its own antiparticle and where $\sigma \times \psi = \sigma$. Ising anyons can implement only the Clifford group by braiding,~\cite{Ahlbrecht09,Bombin10,Hwang24} and, hence, are not universal for quantum computation. The matrix forms of gates for Fibonacci and Ising anyons are shown in panels (B) and (D), respectively, in Fig.~\ref{fig2LS}. The braid groups can be represented in terms of generators, $\sigma_i$, which are counterclockwise exchanges between $i$th and $(i+1)$th anyon particle. A few properties of the braid group generators are shown in Fig.~\ref{figBraidDiagrams}.

Digital quantum simulations on superconducting quantum processors have demonstrated Fibonacci fusion rules and non-Abelian braiding statistics,~\cite{minev25,xu24braid} while Fibonacci anyons are predicted to emerge as quasiparticle excitations in topologically ordered phases.~\cite{Hormozi09} Ising anyons have also been simulated using superconducting qubits, with their fusion rules and non-Abelian braiding statistics experimentally demonstrated.~\cite{Andersen23} Recently, Microsoft reported progress towards a Majorana zero mode based quantum processor, although the topological nature of the reported devices remains under active scientific debate.~\cite{Aghaee25,legg26}

While quantum hardware is in a nascent stage of development, the establishment of algorithms capable of simulating real time quantum dynamics is a critical priority.~\cite{Preskill18} To date, quantum computing applications have largely been confined to the calculation of static molecular properties via hybrid quantum-classical protocols, with only limited progress toward real time dynamics in such hybrid systems.~\cite{cao19,dutta24,cabral24,vu25} We address this limitation by providing a framework for the complete representation of quantum dynamics through unitary operators, enabling fully quantum execution. By leveraging the braiding statistics of both Ising and Fibonacci anyons, we synthesize complex propagators across a rigorous hierarchy of complexity; from fundamental two level systems and one dimensional potentials to many-body condensed phase models and full molecular reaction kinetics. Given the inherent fault-tolerance of topological quantum computers, our methodology establishes a scalable and robust pathway for exploring nonequilibrium phenomena and strongly correlated states in condensed phase chemistry. To our knowledge, this is the first attempt to devise topological quantum algorithms for addressing quantum dynamics in chemical systems.

\section{Solovay-Kitaev Algorithm for Braiding Anyons}\label{secSK}

A central challenge in implementing quantum dynamics on physical hardware is gate synthesis, identifying a sequence of hardware-native gates that accurately approximates a target unitary propagator. Quantum computers efficiently perform unitary matrix multiplication, but available gates are constrained by the device’s physics and braiding statistics of non-Abelian anyons.~\cite{lahtinen17} A brute-force search for these sequences is exponentially time-consuming. To overcome this, we employ the Solovay-Kitaev (SK) algorithm,~\cite{kitaev97,dawson05,NielsenChuangBook} a fundamental result in quantum information theory that approximates arbitrary unitary operations using a finite, universal gate set.

The SK algorithm's primary advantage is its efficiency, ensuring that the sequence length required to reach a target precision scales only polylogarithmically (operating in polynomial time) with that accuracy.~\cite{dawson05} For simulating real time dynamics, the SK algorithm is particularly well suited because it approximates unitaries up to a global phase, which cancels out when computing physical observables like expectation values and correlation functions. While newer techniques offer further optimizations in braid length,~\cite{Xu08,Xu09,Xu11,Burrello10} the SK algorithm provides a rigorous and well-established baseline for synthesizing the propagators required for the hierarchy of systems investigated in this work. A concise mathematical description of the SK procedure is provided in \supp{}.

\section{Dynamics of a Bare Two Level System}

As a first test case, we consider a two-level system with the Hamiltonian $H = \epsilon \sigma_z + J \sigma_x$, where $\sigma_x$ and $\sigma_z$ are Pauli matrices, $2\epsilon$ represents the diabatic energy gap, and $J$ denotes the diabatic coupling between the two states. This system is sufficiently simple that the exact dynamics of the density matrix, $\rho$, can be derived analytically,
\begin{equation}
\rho(t) = e^{-iHt/\hbar}\rho(0)e^{iHt/\hbar},
\end{equation}
where the exact propagator,
\begin{equation}
e^{-iHt/\hbar} = \cos(\Omega t) - iH\sin(\Omega t),
\end{equation}
with $\Omega = \sqrt{J^2 + \epsilon^2}/\hbar$. The propagator is approximated with both Fibonacci and Ising anyons. The accuracy of anyon braiding is contingent upon the target accuracy (or infidelity, as defined in Eq.~\eqref{eqappInfidelity}) achieved through the SK algorithm. For example, to obtain an infidelity of 3.5$\times$10$^{-6}$ with Fibonacci anyons and a time step of $0.2\hbar$, and $\epsilon=0.3,\,J=0.6$, the following braid sequence is needed:
\begin{equation}
e^{-iH\Delta t/\hbar} \approx (\sigma_1^{-1})^2 \sigma_2^{-1} \sigma_1 \sigma_2 \sigma_1^{-1} \sigma_2 \sigma_1 \sigma_2^{-1} \sigma_1 \sigma_2^{-1} \sigma_1^{-1} \sigma_2 \sigma_1.
\end{equation}
Results are shown in Fig.~\ref{fig2LS}. For the remainder of the results presented in this work, Fibonacci anyon braiding has been utilized. An accuracy of $10^{-8}$ has been employed throughout, unless otherwise specified. We predominantly utilized Fibonacci anyons due to the larger basis set of Ising anyons (6 for Ising, 4 for Fibonacci) and the corresponding requirement for significantly larger topological braids.~\cite{Baraban10}

\section{Time correlation Function with Double Well Potential}\label{secDoubleWell}
\begin{figure}
    \centering
    \includegraphics[width=\linewidth]{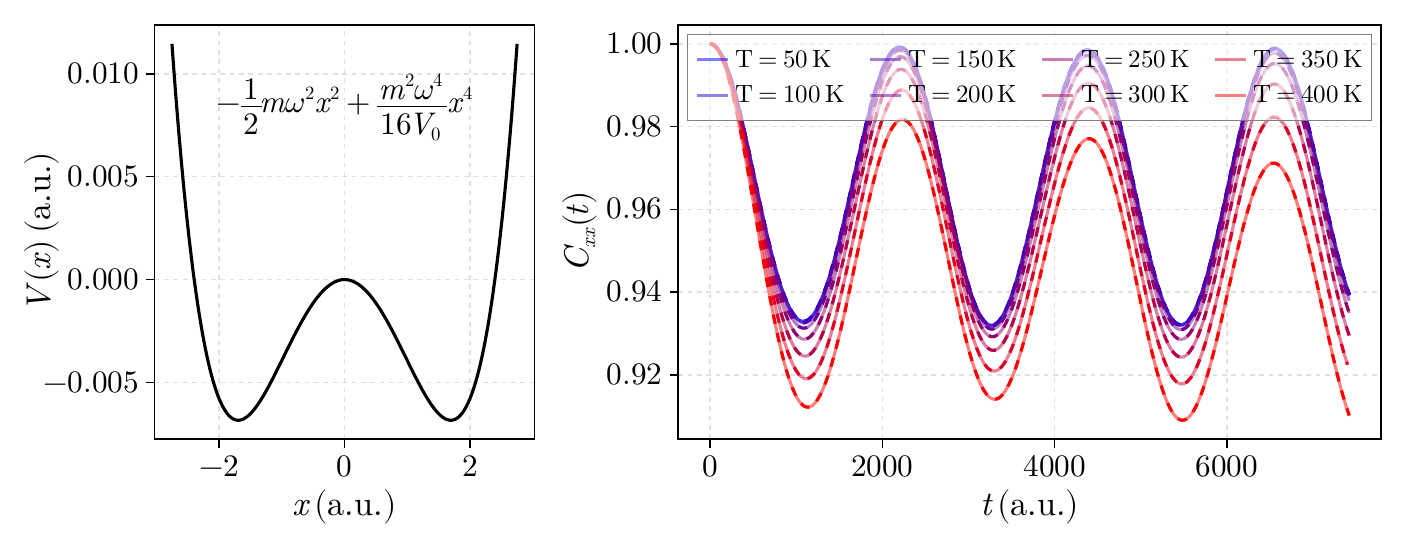}
    \caption{The real part of the position autocorrelation function for a double well potential is displayed on the right. The shape of the potential is shown in the left. We used the parameters from Ref.~\citenum{eklund26}, with $\omega=500\, \mathrm{cm}^{-1}$, $m=1836 \,\mathrm{a.u.}$, and $V_0=1500\,\mathrm{cm}^{-1}$. The position autocorrelation function is shown at different temperatures, with the solid line being the result from topological anyon braiding and dashed line being the exact result.}
    \label{figDoubleWell}
\end{figure}

Next, we test a multilevel system, specifically dynamics in a one-dimensional double well potential, which can be solved using a grid. For the implementation of a multilevel problem (where the size of the matrix exceeds 2), we employ a unitary decomposition using Givens rotations to map a large unitary matrix into a product of numerous effective two-level unitary matrices. Subsequently, we utilize embedding to map the unitary matrices on a quantum circuit.~\cite{NielsenChuangBook,givens58,barenco97,krol22}

The right panel of Fig.~\ref{figDoubleWell} shows the real part of the position autocorrelation function at various temperatures. We compare an exact calculation performed on a 1001 point grid with a braiding approach using a 50 state basis. The unitary propagator ($e^{-iH\Delta t/\hbar}$, where the double well potential is depicted in the left panel of Fig.~\ref{figDoubleWell}) was approximated using Fibonacci anyon braiding. The initial density matrix is sampled from a thermal distribution to incorporate temperature effects. The real part of the position autocorrelation function computed with the braiding algorithm exhibits excellent agreement with the exact result across a range of temperatures. The parameters for the double well potential are taken from Ref.~\citenum{eklund26}. Appendix~\ref{appCost} provides a comment on the number of gates required for this double well potential.

\section{Condensed Phase Systems: Spin Boson Model}
Until now, we have considered small, one-dimensional systems where we could compute the propagator numerically exactly and use the SK algorithm to approximate unitary propagators with desired precision. Next, we move to a condensed phase system, the spin boson model, where the exponential form of the full propagator is not feasible due to many degrees of freedom. The spin-boson model has the Hamiltonian,
\begin{gather}
    H  = H_s + H_b + H_{sb}, \label{eqSBHTot}\\
    H_s = \epsilon \sigma_z + J \sigma_x, \quad H_b = \sum_{j} \hbar\omega_{j} \left(a_j^{\dagger}a_j + \frac{1}{2}\right), \label{eqSBhshb}\\
    H_{sb} =  \sigma_z\sum_{j}\kappa_j (a_j + a_j^{\dagger}), \label{eqSBHsb}
\end{gather}
with $\sigma_{x,z}$ being Pauli matrices, $2\epsilon$ the diabatic energy gap, $J$ the diabatic coupling, and $a_j$ ($a_j^{\dagger}$) are the bosonic annihilation (creation) operators. The system has a bilinear coupling, $\kappa_j$, with a bath of harmonic oscillators. Despite its simplicity, this model cannot be solved analytically and shows various non-Markovian phenomena, including quantum phase transition between localized and delocalized states of the two-level system at finite temperature.~\cite{leggett87}

\begin{figure}
    \centering
    \includegraphics[width=\linewidth]{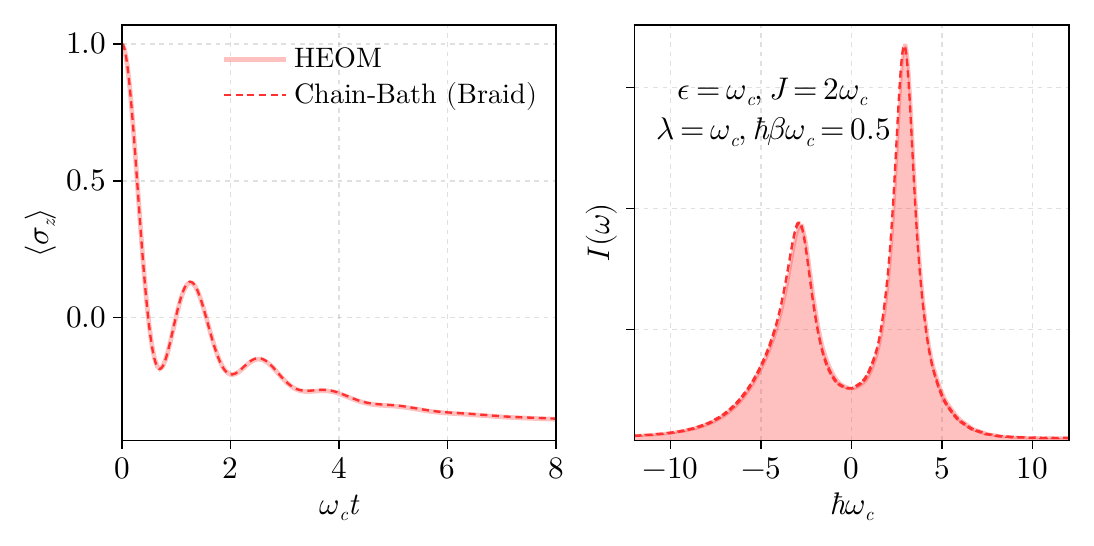}
    \caption{Expectation value of $\sigma_z$ for a spin Boson model is displayed on the left. The parameters are shown on the right panel. The solid light curve is the numerically exact hierarchical equations of motion (HEOM)~\cite{tanimura14,tanimura20} result while the dashed line is the result obtained with topological braiding. A Lorentzian spectral density of the form, $J(\omega) = 2\lambda \omega \omega_c/ (\omega^2+\omega_c^2)$, was utilized to sample the bath frequencies and coupling constants where $\lambda$ is the solvent reorganization energy and $\omega_c$ is the peak frequency of the spectral density. The absorption spectra are plotted on the right, the shaded region being the HEOM result. For computing spectra, we added a dummy ground state (state 0), and used to dipole transitions to the states 1 and 2, $\mu_{01}/\mu_{02}=-5$.}
    \label{figSpinBoson}
\end{figure}

We use a tensor network based path integral method for dynamical propagation of this system.~\cite{chin10,prior10} The details of the propagation scheme with braiding algorithm is provided in \supp{}. The left column shows expectation value of $\sigma_z$ for a spin Boson model and the absorption spectra on the right in Fig.~\ref{figSpinBoson}. The absorption spectra are obtained by taking a Fourier transform of the dipole autocorrelation function.~\cite{mukamelBook}

\section{Molecular Systems: Vibronic Spectra for H\texorpdfstring{$_2$}{H2}}
To treat nuclear and electronic degrees of freedom in a molecular system on equal footing, we use second quantization formulation of fermionic Hamiltonian using spin-orbitals for electronic states, as the Fock states can be encoded on the qubits,~\cite{Ortiz01,delarco25}
\begin{gather}
    H = \sum_{j}\frac{P_j^2}{2M_j} + V_0(\bm{R}) + \sum_{j,k}h_{jk}(\bm{R})c_{j}^{\dagger}c_k \nonumber \\
    + \sum_{j>k} U_{jk}(\bm{R}) c_{j}^{\dagger}c_{j}c_{k}^{\dagger}c_{k} +  \frac{1}{2} \sum_{\substack{j\neq k \,\&\, l\neq n\\ j,k,l,n}} h_{jkln} (\bm{R}) c_{j}^{\dagger}c_k^{\dagger}c_lc_n, \label{eqSecondQHamil}
\end{gather}
where $\bm{P}$ and $\bm{R}$ are nuclear momentum and position, respectively, $c^{\dagger}_{j}$ ($c_{j}$) are fermionic creation (annihilation) operators. $h_{jk}$, $U_{jk}$, and $h_{jkln}$  are the one electron integrals, the Coulomb repulsion energy terms, and two electron exchange terms, respectively. We use Jordan-Wigner transformations~\cite{fradkin89,batista01} to map the system of fermions to a system of qubits.~\cite{Whitfield11} We propagate the dynamics using fermionic path integral, employing Grassmann variables with fermionic coherent states.~\cite{Catto13,xu24} Details of the propagation scheme and path integrals with Grassmann variables are provided in \supp{}. We use this formulation to represent the propagator as a product of unitary matrices, and compute it with topological braiding operations. The vibronic spectrum of H$_2$ is calculated from its transition dipole moment autocorrelation function.

\begin{figure}
    \centering
    \includegraphics[width=0.9\linewidth]{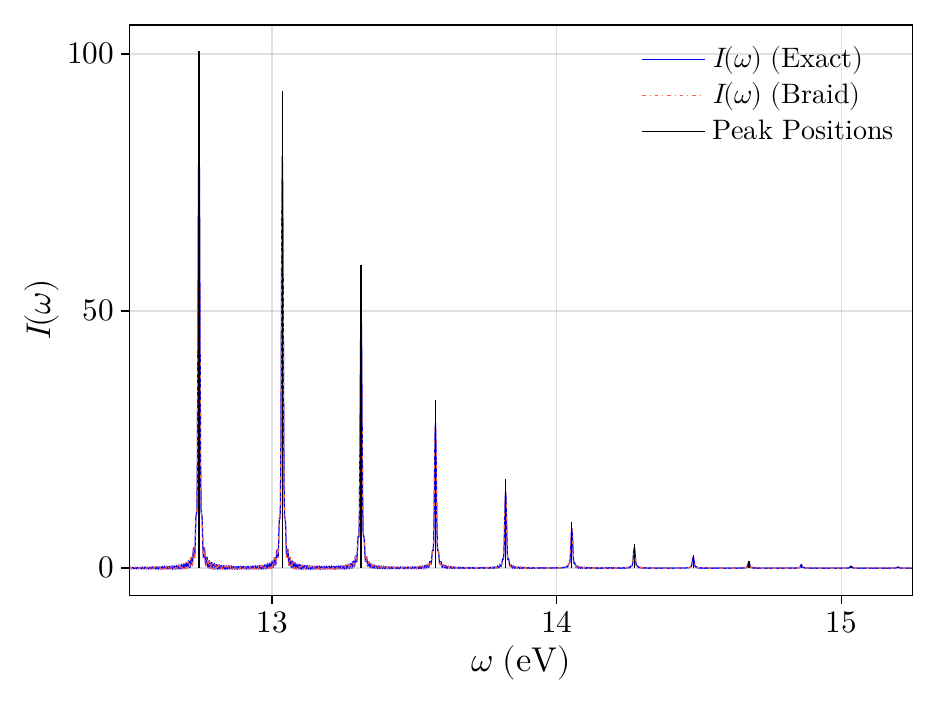}
    \caption{Vibronic spectra for H$_2$ molecule at $T=0\,\mathrm{K}$ for $\sigma\rightarrow\pi$ transition. The position of the peaks of the spectra is shown in the black vertical lines.}
    \label{figSpectraH2}
\end{figure}

The vibronic spectrum for $\sigma\rightarrow\pi$ transition at $0\,\mathrm{K}$ for H$_2$ molecule is shown in Fig.~\ref{figSpectraH2}. The one and two electron integrals in Eq.~\eqref{eqSecondQHamil} are computed with aug-cc-pV5Z basis set~\cite{kendall92} with PySCF library.~\cite{sun20} We use a two-electron/four-orbital model ($\sigma, \sigma^{*}, \pi_x, \pi_y$; total 8 with spin components). The peaks in the spectra show vibronic progression. At $T=0\,\mathrm{K}$ the peaks should ideally be delta functions, and hence a small decay function is added to the time domain dipole autocorrelation to remove noise from the Fourier transform. The peak positions in the computed spectra align correctly with the peak locations obtained from matrix diagonalization. Furthermore, the peak positions also align well with experimental observations.~\cite{Philip04,Bailly10}

\section{Reaction Rate: H + H\texorpdfstring{$_2\rightarrow$}{H2arrow}  H\texorpdfstring{$_2$}{H2} + H}\label{secH2Hreaction}
\begin{figure}
    \centering
    \includegraphics[width=0.95\linewidth]{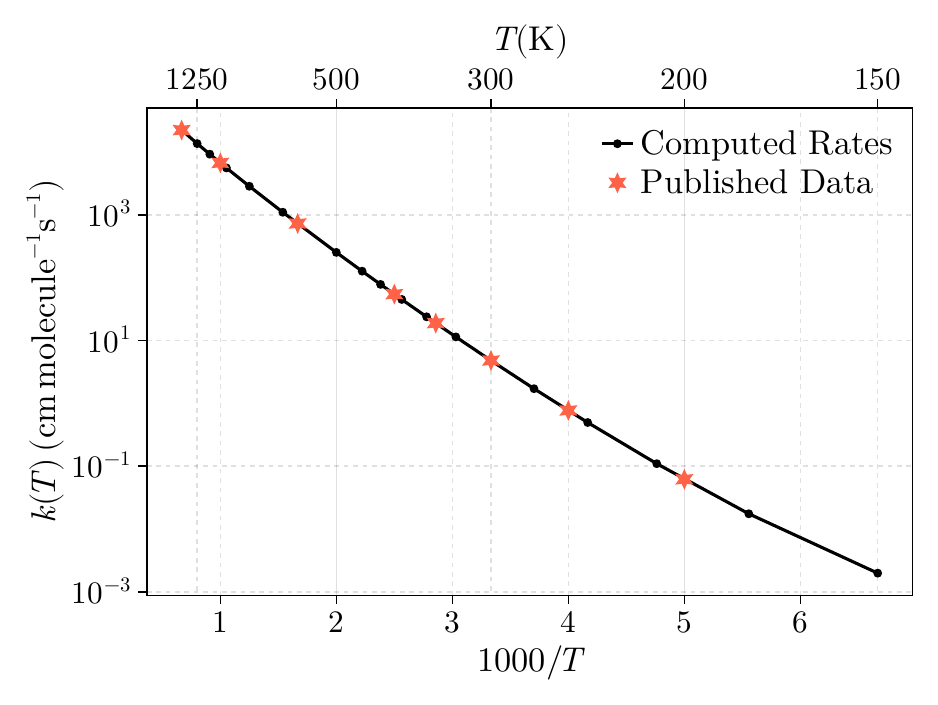}
    \caption{Collinear reaction rate for the reaction H+H$_2 \rightarrow$ H$_2$ + H is displayed above over wide temperature range. The LSTH potential is used  to calculate the flux-flux correlation function using topological braids for unitary operators. The published data, red stars, are taken from Ref.~\citenum{bondi82}.}
    \label{figRateH2H}
\end{figure}

In this last example, we investigate the collinear reaction of H + H$_2 \rightarrow$ H$_2$ + H at various temperatures. We employ a flux-flux correlation function for the 1D reaction rate, as described in Refs.~\citenum{miller83,tromp87}. A Liu-Siegbahn-Truhlar-Horowitz (LSTH) potential for the H$_3$ system is utilized, which has been shown to yield a sufficiently accurate description of the potential energy surface.~\cite{siegbahn78,Truhlar78} The unitary propagators are approximated using topological braids, and the thermal operator $e^{-\beta H}$ can be computed using the probabilistic imaginary time evolution (PITE) method.~\cite{xie24,Nishi23}

PITE achieves unitarity via the use of an ancillary qubit to the quantum state $|\psi\rangle$. The thermal operator, $\mathcal{M}(\tau)=e^{-\tau H}$, where $\tau$ is a small imaginary time step, is expanded into a larger unitary matrix which is approximated using quantum gates,
\begin{equation}
    U_{\mathrm{PITE}}(\tau) = \begin{pmatrix}
        \mathcal{M} & \sqrt{1-\mathcal{M}\mathcal{M}^{\dagger}} \\
        \sqrt{1-\mathcal{M}^{\dagger}\mathcal{M}} & -\mathcal{M}^{\dagger}
    \end{pmatrix}. \label{eqPITEexapnd}
\end{equation}                     
The specific form of the algorithm for short imaginary time step is
\begin{align}
    U_{\rm PITE}(\tau)|\psi\rangle\otimes|0\rangle &  = m_0\left(1-\tau H\right)|\psi\rangle\otimes|0\rangle \nonumber \\
    + & \left(r + \frac{m_0^2 \tau H}{r}\right) |\psi\rangle\otimes|1\rangle + {\mathcal O}(\tau^2),
\end{align}
where $r=\sqrt{1-m_0^2}$, and $m_0$ is an adjustable real parameter such that $0 < m_0 < 1$, $m_0 \neq 1/\sqrt{2}$. Because the approximate non-unitary thermal propagator is associated with the $|0\rangle$ state of the ancilla, the extraction of this operator from the extended matrix, Eq.~\eqref{eqPITEexapnd}, is probabilistic.

The results for reaction rates of H + H$_2 \rightarrow$ H$_2$ + H on the LSTH potential are presented in Fig.~\ref{figRateH2H} and compared against previously published data from Ref.~\citenum{bondi82}. Throughout the temperature range explored, the computed rates with anyon braiding align closely with the previously calculated results by \citeauthor{bondi82}.

\section{Outlook}

In this work, we have established a comprehensive framework for simulating full quantum dynamics on topological quantum hardware, surpassing the current limitations of hybrid quantum-classical hardware methods. By harnessing the fault-tolerant properties of non-Abelian anyons, we demonstrated that complex unitary propagators can be efficiently synthesized using the Solovay-Kitaev algorithm. Our results validate this approach across a rigorous hierarchy of complexity, encompassing fundamental two-level systems and one-dimensional potential wells, as well as many-body condensed phase spin-boson models, culminating in full molecular dynamics for the rate of reaction. By successfully capturing non-Markovian dynamics and vibronic spectra with high fidelity, we have demonstrated that topological braiding will serve as a viable computational vehicle for real time chemical physics. The methodologies presented in this work are agnostic to the nature of quantum hardware, as we present the dynamics in terms of small unitary matrices.

The significance of this framework lies in its ability to provide a scalable pathway for simulating many-body phenomena that are currently inaccessible to classical methods. While current quantum chemistry is often restricted to the calculation of static ground state properties, our methodology paves the way for the exploration of real time dynamics, including non-adiabatic transitions, high-temperature reaction kinetics, and the evolution of strongly correlated electronic states. Future work will focus on extending these algorithms to larger multi-orbital molecular systems and more complex, realistic bath environments, and fermionic molecules in a fermionic environment.~\cite{farajollahpour26,lee21} Ultimately, this work provides a foundational methodology for utilizing topological quantum computation to solve some of the most challenging and chemically relevant problems in condensed phase physics.

\vspace{0.1in}
\noindent
\textbf{Data Availability:} Codes for reproducing the figures in this manuscript are available at \href{https://github.com/kritanjan-polley/Topological-Quantum-Computing-with-Anyons.git}{GitHub}. \supp{} is available.

\vspace{0.1in}
\noindent
\textbf{Acknowledgments:} This work was supported by a grant from the Simons Foundation [MPS-T-MPS-00839534, MET]. M.E.T. acknowledges support from NSF grant \#2536174 AMD 000.

\bibliography{reference}

@article{friedman77,
author = {Friedman, Jerome H. and Bentley, Jon Louis and Finkel, Raphael Ari},
title = {An Algorithm for Finding Best Matches in Logarithmic Expected Time},
year = {1977},
issue_date = {Sept. 1977},
publisher = {Association for Computing Machinery},
address = {New York, NY, USA},
volume = {3},
number = {3},
issn = {0098-3500},
url = {https://doi.org/10.1145/355744.355745},
doi = {10.1145/355744.355745},
journal = {ACM Trans. Math. Softw.},
month = sep,
pages = {209–226},
numpages = {18}
}

@article{Nishi23,
  title = {Optimal scheduling in probabilistic imaginary-time evolution on a quantum computer},
  author = {Nishi, Hirofumi and Hamada, Koki and Nishiya, Yusuke and Kosugi, Taichi and Matsushita, Yu-ichiro},
  journal = {Phys. Rev. Res.},
  volume = {5},
  issue = {4},
  pages = {043048},
  numpages = {14},
  year = {2023},
  month = {Oct},
  publisher = {American Physical Society},
  doi = {10.1103/PhysRevResearch.5.043048},
  url = {https://link.aps.org/doi/10.1103/PhysRevResearch.5.043048}
}

@article{Philip04,
author = {Philip, J and Sprengers, J P and Pielage, Th. and de Lange, C A and Ubachs, W and Reinhold, E},
title = {Highly accurate transition frequencies in the H2 Lyman and Werner absorption bands},
journal = {Canadian Journal of Chemistry},
volume = {82},
number = {6},
pages = {713-722},
year = {2004},
doi = {10.1139/v04-042},
URL = {https://doi.org/10.1139/v04-042},
}

@article{Bailly10,
author = {D. Bailly and E.J. Salumbides and M. Vervloet and W. Ubachs},
title = {Accurate level energies in the EF1 , GK1 , H1 , B1 , , B′1 , , , J1Δg states of H2 },
journal = {Molecular Physics},
volume = {108},
number = {7-9},
pages = {827--846},
year = {2010},
publisher = {Taylor \& Francis},
doi = {10.1080/00268970903413350},
URL = {https://doi.org/10.1080/00268970903413350},
}

@article{legg26,
  title={On the robustness of topological gap detection via transport},
  author={Legg, Henry F},
  journal={Nature},
  volume={654},
  number={8120},
  pages={E22--E26},
  year={2026},
  doi={10.1038/s41586-026-10567-8},
  publisher={Nature Publishing Group UK London}
}

@Article{Aghaee25,
author="Aghaee, Morteza
and Alcaraz Ramirez, Alejandro
and Alam, Zulfi
and Ali, Rizwan
and Andrzejczuk, Mariusz
and Antipov, Andrey
and Astafev, Mikhail
and Barzegar, Amin
and Bauer, Bela
and Becker, Jonathan
and others",
title="Interferometric single-shot parity measurement in InAs--Al hybrid devices",
journal="Nature",
year="2025",
month="Feb",
day="01",
volume="638",
number="8051",
pages="651--655",
issn="1476-4687",
doi="10.1038/s41586-024-08445-2",
url="https://doi.org/10.1038/s41586-024-08445-2"
}

@Article{Andersen23,
author="Andersen, T. I.
and Lensky, Y. D.
and Kechedzhi, K.
and Drozdov, I. K.
and Bengtsson, A.
and Hong, S.
and Morvan, A.
and Mi, X.
and Opremcak, A.
and others",
title="Non-Abelian braiding of graph vertices in a superconducting processor",
journal="Nature",
year="2023",
month="Jun",
day="01",
volume="618",
number="7964",
pages="264--269",
issn="1476-4687",
doi="10.1038/s41586-023-05954-4",
url="https://doi.org/10.1038/s41586-023-05954-4"
}

@article{Hormozi09,
  title = {Topological Quantum Computing with Read-Rezayi States},
  author = {Hormozi, L. and Bonesteel, N. E. and Simon, S. H.},
  journal = {Phys. Rev. Lett.},
  volume = {103},
  issue = {16},
  pages = {160501},
  numpages = {4},
  year = {2009},
  month = {Oct},
  publisher = {American Physical Society},
  doi = {10.1103/PhysRevLett.103.160501},
  url = {https://link.aps.org/doi/10.1103/PhysRevLett.103.160501}
}

@article{Ortiz01,
  title = {Quantum algorithms for fermionic simulations},
  author = {Ortiz, G. and Gubernatis, J. E. and Knill, E. and Laflamme, R.},
  journal = {Phys. Rev. A},
  volume = {64},
  issue = {2},
  pages = {022319},
  numpages = {14},
  year = {2001},
  month = {Jul},
  publisher = {American Physical Society},
  doi = {10.1103/PhysRevA.64.022319},
  url = {https://link.aps.org/doi/10.1103/PhysRevA.64.022319}
}

@misc{ECC,
  doi = {10.48550/ARXIV.2508.01879},
  url = {https://arxiv.org/abs/2508.01879},
  author = {Tham,  Edwin and Ye,  Min and Khait,  Ilia and Gamble,  John and Delfosse,  Nicolas},
  title = {Distributed fault-tolerant quantum memories over a 2xL array of qubit modules},
  publisher = {arXiv},
  year = {2025},
  copyright = {Creative Commons Attribution 4.0 International}
}

@article{Baraban10,
  title = {Resources required for topological quantum factoring},
  author = {Baraban, M. and Bonesteel, N. E. and Simon, S. H.},
  journal = {Phys. Rev. A},
  volume = {81},
  issue = {6},
  pages = {062317},
  numpages = {4},
  year = {2010},
  month = {Jun},
  publisher = {American Physical Society},
  doi = {10.1103/PhysRevA.81.062317},
  url = {https://link.aps.org/doi/10.1103/PhysRevA.81.062317}
}

@misc{farajollahpour26,
      title={Quantum Algorithm Software for Condensed Matter Physics}, 
      author={T. Farajollahpour},
      year={2026},
      eprint={2506.09308},
      archivePrefix={arXiv},
      primaryClass={cond-mat.str-el},
      url={https://arxiv.org/abs/2506.09308}, 
}

@article{lee21,
    author = {Lee, Chee-Kong and Hsieh, Chang-Yu and Zhang, Shengyu and Shi, Liang},
    title = {Simulation of Condensed-Phase Spectroscopy with Near-Term Digital Quantum Computers},
    journal = {J. Chem. Theory Comput.},
    volume = {17},
    number = {11},
    pages = {7178-7186},
    year = {2021},
    month = {10},
    issn = {1549-9618},
    doi = {10.1021/acs.jctc.1c00849},
    url = {https://doi.org/10.1021/acs.jctc.1c00849},
}

@article{tanimura14,
    author = {Tanimura, Yoshitaka},
    title = {Reduced hierarchical equations of motion in real and imaginary time: Correlated initial states and thermodynamic quantities},
    journal = {J. Chem. Phys.},
    volume = {141},
    number = {4},
    pages = {044114},
    year = {2014},
    month = {07},
    issn = {0021-9606},
    doi = {10.1063/1.4890441},
    url = {https://doi.org/10.1063/1.4890441},
}

@article{tanimura20,
    author = {Tanimura, Yoshitaka},
    title = {Numerically “exact” approach to open quantum dynamics: The hierarchical equations of motion (HEOM)},
    journal = {J. Chem. Phys.},
    volume = {153},
    number = {2},
    pages = {020901},
    year = {2020},
    month = {07},
    issn = {0021-9606},
    doi = {10.1063/5.0011599},
    url = {https://doi.org/10.1063/5.0011599},
}

@article{vu25,
    author = {Vu, Nam P. and Dong, Daniel and Dan, Xiaohan and Lyu, Ningyi and Batista, Victor and Liu, Yuan},
    title = {A Computational Framework for Simulations of Dissipative Nonadiabatic Dynamics on Hybrid Oscillator-Qubit Quantum Devices},
    journal = {J. Chem. Theory Comput. },
    volume = {21},
    number = {13},
    pages = {6258-6279},
    year = {2025},
    month = {06},
    issn = {1549-9618},
    doi = {10.1021/acs.jctc.5c00315},
    url = {https://doi.org/10.1021/acs.jctc.5c00315},
}

@article{cabral24,
    author = {Cabral, Delmar G. A. and Khazaei, Pouya and Allen, Brandon C. and Videla, Pablo E. and Schäfer, Max and Cortiñas, Rodrigo G. and Carrillo de Albornoz, Alejandro Cros and Chávez-Carlos, Jorge and Santos, Lea F. and Geva, Eitan and Batista, Victor S.},
    title = {A Roadmap for Simulating Chemical Dynamics on a Parametrically Driven Bosonic Quantum Device},
    journal = {J. Phys. Chem. Lett.},
    volume = {15},
    number = {48},
    pages = {12042-12050},
    year = {2024},
    month = {11},
    issn = {1948-7185},
    doi = {10.1021/acs.jpclett.4c02864},
    url = {https://doi.org/10.1021/acs.jpclett.4c02864},
}

@article{dutta24,
    author = {Dutta, Rishab and Cabral, Delmar G. A. and Lyu, Ningyi and Vu, Nam P. and Wang, Yuchen and Allen, Brandon and Dan, Xiaohan and Cortiñas, Rodrigo G. and Khazaei, Pouya and Schäfer, Max and others},
    title = {Simulating Chemistry on Bosonic Quantum Devices},
    journal = {J. Chem. Theory Comput.},
    volume = {20},
    number = {15},
    pages = {6426-6441},
    year = {2024},
    month = {07},
    issn = {1549-9618},
    doi = {10.1021/acs.jctc.4c00544},
    url = {https://doi.org/10.1021/acs.jctc.4c00544},
}

@article{lahtinen17,
	title = {A Short Introduction to Topological Quantum Computation},
	pages = {021},
	author = {Lahtinen, Ville and Pachos, Jiannis},
	journal = {SciPost Phys.},
	volume = {3},
	year = {2017},
	publisher = {SciPost},
	doi = {10.21468/SciPostPhys.3.3.021},
	url = {https://scipost.org/10.21468/SciPostPhys.3.3.021}
}

@article{cao19,
    author = {Cao, Yudong and Romero, Jonathan and Olson, Jonathan P. and Degroote, Matthias and Johnson, Peter D. and Kieferová, Mária and Kivlichan, Ian D. and Menke, Tim and Peropadre, Borja and Sawaya, Nicolas P. D. and Sim, Sukin and Veis, Libor and Aspuru-Guzik, Alán},
    title = {Quantum Chemistry in the Age of Quantum Computing},
    journal = {Chem. Rev.},
    volume = {119},
    number = {19},
    pages = {10856-10915},
    year = {2019},
    month = {08},
    issn = {0009-2665},
    doi = {10.1021/acs.chemrev.8b00803},
    url = {https://doi.org/10.1021/acs.chemrev.8b00803},
}

@article{Preskill18,
  doi = {10.22331/q-2018-08-06-79},
  url = {https://doi.org/10.22331/q-2018-08-06-79},
  title = {Quantum {C}omputing in the {NISQ} era and beyond},
  author = {Preskill, John},
  journal = {{Quantum}},
  issn = {2521-327X},
  publisher = {{Verein zur F{\"{o}}rderung des Open Access Publizierens in den Quantenwissenschaften}},
  volume = {2},
  pages = {79},
  month = aug,
  year = {2018}
}

@article{xu24braid,
  title={Non-Abelian braiding of Fibonacci anyons with a superconducting processor},
  author={Xu, Shibo and Sun, Zheng-Zhi and Wang, Ke and Li, Hekang and Zhu, Zitian and Dong, Hang and Deng, Jinfeng and Zhang, Xu and Chen, Jiachen and Wu, Yaozu and others},
  journal={Nat. Phys.},
  volume={20},
  number={9},
  pages={1469--1475},
  year={2024},
  doi={10.1038/s41567-024-02529-6},
  publisher={Nature Publishing Group UK London}
}

@article{xue26,
  title = {Statistics of Abelian topological excitations},
  author = {Xue, Hanyu},
  journal = {Phys. Rev. B},
  volume = {113},
  issue = {4},
  pages = {045143},
  numpages = {36},
  year = {2026},
  month = {Jan},
  publisher = {American Physical Society},
  doi = {10.1103/g3nc-fwqg},
  url = {https://link.aps.org/doi/10.1103/g3nc-fwqg}
}

@article{lutchyn18,
  title={Majorana zero modes in superconductor--semiconductor heterostructures},
  author={Lutchyn, Roman M and Bakkers, Erik PAM and Kouwenhoven, Leo P and Krogstrup, Peter and Marcus, Charles M and Oreg, Yuval},
  journal={Nat. Rev. Mater.},
  volume={3},
  number={5},
  pages={52--68},
  year={2018},
  doi={10.1038/s41578-018-0003-1},
  publisher={Nature Publishing Group UK London}
}

@article{sarma15,
  title={Majorana zero modes and topological quantum computation},
  author={Sarma, Sankar Das and Freedman, Michael and Nayak, Chetan},
  journal={npj Quantum Inf.},
  volume={1},
  number={1},
  pages={15001},
  year={2015},
  doi={10.1038/npjqi.2015.1},
  publisher={Nature Publishing Group}
}

@book{StanescuBook,
author = {Stanescu, Tudor D.},
address = {Boca Raton},
edition = {Second edition.},
isbn = {9781040041987},
publisher = {CRC Press, Taylor \& Francis Group},
title = {Introduction to topological quantum matter \& quantum computation },
year = {2024 - 2025},
doi={10.1201/9781003226048},
}

@article{daskin11,
    author = {Daskin, Anmer and Kais, Sabre},
    title = {Decomposition of unitary matrices for finding quantum circuits: Application to molecular Hamiltonians},
    journal = {J. Chem. Phys.},
    volume = {134},
    number = {14},
    pages = {144112},
    year = {2011},
    month = {04},
    issn = {0021-9606},
    doi = {10.1063/1.3575402},
    url = {https://doi.org/10.1063/1.3575402},
}

@article{tacchino20,
author = {Tacchino, Francesco and Chiesa, Alessandro and Carretta, Stefano and Gerace, Dario},
title = {Quantum Computers as Universal Quantum Simulators: State-of-the-Art and Perspectives},
journal = {Advanced Quantum Technologies},
volume = {3},
number = {3},
pages = {1900052},
doi = {https://doi.org/10.1002/qute.201900052},
url = {https://advanced.onlinelibrary.wiley.com/doi/abs/10.1002/qute.201900052},
year = {2020}
}

@book{Mermin_2007, 
place={Cambridge}, 
title={Quantum Computer Science: An Introduction}, 
publisher={Cambridge University Press}, 
author={Mermin, N. David}, 
doi={10.1017/CBO9780511813870},
year={2007}
}

@article{Aaronson08,
 ISSN = {00368733, 19467087},
 URL = {http://www.jstor.org/stable/26000518},
 author = {Scott Aaronson},
 journal = {Scientific American},
 number = {3},
 pages = {62--69},
 publisher = {Scientific American, a division of Nature America, Inc.},
 title = {THE LIMITS OF Quantum},
 urldate = {2026-08-16},
 volume = {298},
 year = {2008}
}

@misc{martonosi19,
      title={Next Steps in Quantum Computing: Computer Science's Role}, 
      author={Margaret Martonosi and Martin Roetteler},
      year={2019},
      eprint={1903.10541},
      archivePrefix={arXiv},
      url={https://arxiv.org/abs/1903.10541}, 
}

@article{Bonderson08,
title = {Interferometry of non-Abelian anyons},
journal = {Annals of Physics},
volume = {323},
number = {11},
pages = {2709-2755},
year = {2008},
issn = {0003-4916},
doi = {https://doi.org/10.1016/j.aop.2008.01.012},
url = {https://www.sciencedirect.com/science/article/pii/S0003491608000171},
author = {Parsa Bonderson and Kirill Shtengel and J.K. Slingerland},
}

@article{Bombin10,
  title = {Topological Order with a Twist: Ising Anyons from an Abelian Model},
  author = {Bombin, H.},
  journal = {Phys. Rev. Lett.},
  volume = {105},
  issue = {3},
  pages = {030403},
  numpages = {4},
  year = {2010},
  month = {Jul},
  publisher = {American Physical Society},
  doi = {10.1103/PhysRevLett.105.030403},
  url = {https://link.aps.org/doi/10.1103/PhysRevLett.105.030403}
}

@article{Hwang24,
  title = {Anyon condensation and confinement transition in a Kitaev spin liquid bilayer},
  author = {Hwang, Kyusung},
  journal = {Phys. Rev. B},
  volume = {109},
  issue = {13},
  pages = {134412},
  numpages = {18},
  year = {2024},
  month = {Apr},
  publisher = {American Physical Society},
  doi = {10.1103/PhysRevB.109.134412},
  url = {https://link.aps.org/doi/10.1103/PhysRevB.109.134412}
}

@article{Ahlbrecht09,
  title = {Implementation of Clifford gates in the Ising-anyon topological quantum computer},
  author = {Ahlbrecht, Andr\'e and Georgiev, Lachezar S. and Werner, Reinhard F.},
  journal = {Phys. Rev. A},
  volume = {79},
  issue = {3},
  pages = {032311},
  numpages = {16},
  year = {2009},
  month = {Mar},
  publisher = {American Physical Society},
  doi = {10.1103/PhysRevA.79.032311},
  url = {https://link.aps.org/doi/10.1103/PhysRevA.79.032311}
}

@article{trebst08,
    author = {Trebst, Simon and Troyer, Matthias and Wang, Zhenghan and Ludwig, Andreas W. W.},
    title = {A Short Introduction to Fibonacci Anyon Models},
    journal = {Prog. Theor. Phys. Suppl. },
    volume = {176},
    pages = {384-407},
    year = {2008},
    month = {06},
    issn = {0375-9687},
    doi = {10.1143/PTPS.176.384},
    url = {https://doi.org/10.1143/PTPS.176.384},
}

@article{Xu08,
  title = {Constructing functional braids for low-leakage topological quantum computing},
  author = {Xu, Haitan and Wan, Xin},
  journal = {Phys. Rev. A},
  volume = {78},
  issue = {4},
  pages = {042325},
  numpages = {4},
  year = {2008},
  month = {Oct},
  publisher = {American Physical Society},
  doi = {10.1103/PhysRevA.78.042325},
  url = {https://link.aps.org/doi/10.1103/PhysRevA.78.042325}
}

@article{Xu09,
  title = {Exploiting geometric degrees of freedom in topological quantum computing},
  author = {Xu, Haitan and Wan, Xin},
  journal = {Phys. Rev. A},
  volume = {80},
  issue = {1},
  pages = {012306},
  numpages = {5},
  year = {2009},
  month = {Jul},
  publisher = {American Physical Society},
  doi = {10.1103/PhysRevA.80.012306},
  url = {https://link.aps.org/doi/10.1103/PhysRevA.80.012306}
}

@article{Xu11,
  title = {Unified approach to topological quantum computation with anyons: From qubit encoding to Toffoli gate},
  author = {Xu, Haitan and Taylor, J. M.},
  journal = {Phys. Rev. A},
  volume = {84},
  issue = {1},
  pages = {012332},
  numpages = {4},
  year = {2011},
  month = {Jul},
  publisher = {American Physical Society},
  doi = {10.1103/PhysRevA.84.012332},
  url = {https://link.aps.org/doi/10.1103/PhysRevA.84.012332}
}

@article{Burrello10,
  title = {Topological Quantum Hashing with the Icosahedral Group},
  author = {Burrello, Michele and Xu, Haitan and Mussardo, Giuseppe and Wan, Xin},
  journal = {Phys. Rev. Lett.},
  volume = {104},
  issue = {16},
  pages = {160502},
  numpages = {4},
  year = {2010},
  month = {Apr},
  publisher = {American Physical Society},
  doi = {10.1103/PhysRevLett.104.160502},
  url = {https://link.aps.org/doi/10.1103/PhysRevLett.104.160502}
}

@Article{minev25,
author="Minev, Zlatko K. and Najafi, Khadijeh and Majumder, Swarnadeep and Wang, Juven and Stern, Ady and Kim, Eun-Ah and Jian, Chao-Ming and Zhu, Guanyu",
title="Realizing string-net condensation: Fibonacci anyon braiding for universal gates and sampling chromatic polynomials",
journal="Nat. Commun.",
year="2025",
month="Jul",
day="06",
volume="16",
number="1",
pages="6225",
issn="2041-1723",
doi="10.1038/s41467-025-61493-8",
url="https://doi.org/10.1038/s41467-025-61493-8"
}

@article{haldane17,
  title={Nobel lecture: Topological quantum matter},
  author={Haldane, F Duncan M},
  journal={Rev. Mod. Phys},
  volume={89},
  number={4},
  pages={040502},
  doi={10.1103/RevModPhys.89.040502},
  year={2017}
}

@article{kitaev03,
title = {Fault-tolerant quantum computation by anyons},
journal = {Ann. Phys.},
volume = {303},
number = {1},
pages = {2-30},
year = {2003},
issn = {0003-4916},
doi = {https://doi.org/10.1016/S0003-4916(02)00018-0},
url = {https://www.sciencedirect.com/science/article/pii/S0003491602000180},
author = {A.Yu. Kitaev},
}

@Article{Feynman82,
author="Feynman, Richard P.",
title="Simulating physics with computers",
journal="Int. J. Theor. Phys.",
year="1982",
month="Jun",
day="01",
volume="21",
number="6",
pages="467--488",
issn="1572-9575",
doi="10.1007/BF02650179",
url="https://doi.org/10.1007/BF02650179"
}

@article{das06,
  title={Topological quantum computation},
  author={Das Sarma, Sankar and Freedman, Michael and Nayak, Chetan},
  journal={Physics today},
  volume={59},
  number={7},
  pages={32--38},
  year={2006},
  doi={10.1063/1.2337825},
  publisher={American Institute of Physics}
}

@article{stern13,
author = {Ady Stern  and Netanel H. Lindner },
title = {Topological Quantum Computation—From Basic Concepts to First Experiments},
journal = {Science},
volume = {339},
number = {6124},
pages = {1179-1184},
year = {2013},
doi = {10.1126/science.1231473},
URL = {https://www.science.org/doi/abs/10.1126/science.1231473},
}

@article{freedman03,
  title={Topological quantum computation},
  author={Freedman, Michael and Kitaev, Alexei and Larsen, Michael and Wang, Zhenghan},
  journal={Bull. Amer. Math. Soc. },
  volume={40},
  number={1},
  pages={31--38},
  doi={10.1090/S0273-0979-02-00964-3},
  year={2003}
}

@article{nayak08,
  title = {Non-Abelian anyons and topological quantum computation},
  author = {Nayak, Chetan and Simon, Steven H. and Stern, Ady and Freedman, Michael and Das Sarma, Sankar},
  journal = {Rev. Mod. Phys.},
  volume = {80},
  issue = {3},
  pages = {1083--1159},
  numpages = {0},
  year = {2008},
  month = {Sep},
  publisher = {American Physical Society},
  doi = {10.1103/RevModPhys.80.1083},
  url = {https://link.aps.org/doi/10.1103/RevModPhys.80.1083}
}

@book{peskinBook,
author = {Peskin, Michael Edward and Schroeder, Daniel V.},
address = {New York},
isbn = {9780813345437},
publisher = {Westview Press},
series = {Frontiers in Physics},
title = {An Introduction To Quantum Field Theory.},
doi={10.1201/9780429503559},
year = {1995},
}

@book{mukamelBook,
author = {Mukamel, S.},
address = {New York},
booktitle = {Principles of nonlinear optical spectroscopy},
isbn = {0195092783},
lccn = {94010792},
publisher = {Oxford University Press},
series = {Oxford series in optical and imaging sciences ; 6},
title = {Principles of nonlinear optical spectroscopy },
year = {1995 - 1995},
}

@article{sun20,
    author = {Sun, Qiming and Zhang, Xing and Banerjee, Samragni and Bao, Peng and Barbry, Marc and Blunt, Nick S. and Bogdanov, Nikolay A. and Booth, George H. and Chen, Jia and Cui, Zhi-Hao and Eriksen, Janus J. and Gao, Yang and Guo, Sheng and Hermann, Jan and Hermes, Matthew R. and others},
    title = {Recent developments in the PySCF program package},
    journal = {J. Chem. Phys.},
    volume = {153},
    number = {2},
    pages = {024109},
    year = {2020},
    month = {07},
    issn = {0021-9606},
    doi = {10.1063/5.0006074},
    url = {https://doi.org/10.1063/5.0006074},
}

@article{kendall92,
    author = {Kendall, Rick A. and Dunning, Thom H., Jr. and Harrison, Robert J.},
    title = {Electron affinities of the first‐row atoms revisited. Systematic basis sets and wave functions},
    journal = {J. Chem. Phys.},
    volume = {96},
    number = {9},
    pages = {6796-6806},
    year = {1992},
    month = {05},
    issn = {0021-9606},
    doi = {10.1063/1.462569},
    url = {https://doi.org/10.1063/1.462569},
}

@article{Baxter72,
title = {Partition function of the Eight-Vertex lattice model},
journal = {Annals of Physics},
volume = {70},
number = {1},
pages = {193-228},
year = {1972},
issn = {0003-4916},
doi = {https://doi.org/10.1016/0003-4916(72)90335-1},
url = {https://www.sciencedirect.com/science/article/pii/0003491672903351},
author = {Rodney J Baxter},
}

@article{Yang67,
  title = {Some Exact Results for the Many-Body Problem in one Dimension with Repulsive Delta-Function Interaction},
  author = {Yang, C. N.},
  journal = {Phys. Rev. Lett.},
  volume = {19},
  issue = {23},
  pages = {1312--1315},
  numpages = {0},
  year = {1967},
  month = {Dec},
  publisher = {American Physical Society},
  doi = {10.1103/PhysRevLett.19.1312},
  url = {https://link.aps.org/doi/10.1103/PhysRevLett.19.1312}
}

@article{xie24,
  title = {Probabilistic imaginary-time evolution algorithm based on nonunitary quantum circuits},
  author = {Xie, Hao-Nan and Wei, Shi-Jie and Yang, Fan and Wang, Zheng-An and Chen, Chi-Tong and Fan, Heng and Long, Gui-Lu},
  journal = {Phys. Rev. A},
  volume = {109},
  issue = {5},
  pages = {052414},
  numpages = {15},
  year = {2024},
  month = {May},
  publisher = {American Physical Society},
  doi = {10.1103/PhysRevA.109.052414},
  url = {https://link.aps.org/doi/10.1103/PhysRevA.109.052414}
}

@article{Truhlar78,
    author = {Truhlar, Donald G. and Horowitz, Charles J.},
    title = {Functional representation of Liu and Siegbahn’s accurate ab initio potential energy calculations for H+H2},
    journal = {J. Chem. Phys.},
    volume = {68},
    number = {5},
    pages = {2466-2476},
    year = {1978},
    month = {03},
    issn = {0021-9606},
    doi = {10.1063/1.436019},
    url = {https://doi.org/10.1063/1.436019},
}

@article{siegbahn78,
    author = {Siegbahn, P. and Liu, B.},
    title = {An accurate three‐dimensional potential energy surface for H3},
    journal = {J. Chem. Phys.},
    volume = {68},
    number = {5},
    pages = {2457-2465},
    year = {1978},
    month = {03},
    issn = {0021-9606},
    doi = {10.1063/1.436018},
    url = {https://doi.org/10.1063/1.436018},
}

@article{tromp87,
    author = {Tromp, John W. and Miller, William H.},
    title = {The reactive flux correlation function for collinear reactions H + H2, Cl + HCl and F + H2},
    journal = {Faraday Discuss. Chem. Soc.},
    volume = {84},
    pages = {441-453},
    year = {1987},
    month = {01},
    issn = {0301-7249},
    doi = {10.1039/DC9878400441},
    url = {https://doi.org/10.1039/DC9878400441},
}

@article{miller83,
    author = {Miller, William H. and Schwartz, Steven D. and Tromp, John W.},
    title = {Quantum mechanical rate constants for bimolecular reactions},
    journal = {J. Chem. Phys.},
    volume = {79},
    number = {10},
    pages = {4889-4898},
    year = {1983},
    month = {11},
    issn = {0021-9606},
    doi = {10.1063/1.445581},
    url = {https://doi.org/10.1063/1.445581},
}

@article{bondi82,
    author = {Bondi, D. K. and Clary, D. C. and Connor, J. N. L. and Garrett, Bruce C. and Truhlar, Donald G.},
    title = {Kinetic isotope effects in the Mu+H2 and Mu+D2 reactions: Accurate quantum calculations for the collinear reactions and variational transition state theory predictions for one and three dimensions},
    journal = {J. Chem. Phys.},
    volume = {76},
    number = {10},
    pages = {4986-4995},
    year = {1982},
    month = {05},
    issn = {0021-9606},
    doi = {10.1063/1.442845},
    url = {https://doi.org/10.1063/1.442845},
}

@misc{dawson05,
      title={The Solovay-Kitaev algorithm}, 
      author={Christopher M. Dawson and Michael A. Nielsen},
      year={2005},
      eprint={quant-ph/0505030},
      archivePrefix={arXiv},
      primaryClass={quant-ph},
      url={https://arxiv.org/abs/quant-ph/0505030}, 
}

@book{NielsenChuangBook, 
place={Cambridge}, 
title={Quantum Computation and Quantum Information: 10th Anniversary Edition}, 
publisher={Cambridge University Press}, 
author={Nielsen, Michael A. and Chuang, Isaac L.}, 
doi={10.1017/CBO9780511976667},
year={2010}
}

@article{kitaev97,
doi = {10.1070/RM1997v052n06ABEH002155},
url = {https://doi.org/10.1070/RM1997v052n06ABEH002155},
year = {1997},
month = {dec},
publisher = {},
volume = {52},
number = {6},
pages = {1191},
author = {A Yu Kitaev},
title = {Quantum Computations: Algorithms and Error Correction},
journal = {Russ. Math. Surv.},
}

@article{eklund26,
    author = {Eklund, Elliot C. and Ananth, Nandini},
    title = {Hybrid quantum algorithm for simulating real-time thermal correlation functions},
    journal = {Digital Discovery},
    volume = {5},
    number = {6},
    pages = {2759-2769},
    year = {2026},
    month = {06},
    issn = {2635-098X},
    doi = {10.1039/d5dd00381d}
}

@article{barenco97,
  title = {Elementary gates for quantum computation},
  author = {Barenco, Adriano and Bennett, Charles H. and Cleve, Richard and DiVincenzo, David P. and Margolus, Norman and Shor, Peter and Sleator, Tycho and Smolin, John A. and Weinfurter, Harald},
  journal = {Phys. Rev. A},
  volume = {52},
  issue = {5},
  pages = {3457--3467},
  numpages = {0},
  year = {1995},
  month = {Nov},
  publisher = {American Physical Society},
  doi = {10.1103/PhysRevA.52.3457},
  url = {https://link.aps.org/doi/10.1103/PhysRevA.52.3457}
}

@book{Altland_Simons_book, 
    place={Cambridge}, 
    edition={2}, 
    title={Condensed Matter Field Theory}, 
    publisher={Cambridge University Press}, 
    author={Altland, Alexander and Simons, Ben D.}, 
    year={2010},
    doi={10.1017/CBO9780511789984},
}

@article{givens58,
 ISSN = {03684245},
 URL = {http://www.jstor.org/stable/2098861},
 author = {Wallace Givens},
 journal = {J. Soc. Indust. Appl. Math.},
 number = {1},
 pages = {26--50},
 publisher = {Society for Industrial and Applied Mathematics},
 title = {Computation of Plane Unitary Rotations Transforming a General Matrix to Triangular Form},
 volume = {6},
 year = {1958}
}

@article{krol22,
  title={Efficient decomposition of unitary matrices in quantum circuit compilers},
  author={Krol, Anna M and Sarkar, Aritra and Ashraf, Imran and Al-Ars, Zaid and Bertels, Koen},
  journal={Appl. Sci.},
  volume={12},
  number={2},
  pages={759},
  year={2022},
  url={https://doi.org/10.3390/app12020759},
  publisher={MDPI}
}

@article{leggett87,
  title = {Dynamics of the dissipative two-state system},
  author = {Leggett, A. J. and Chakravarty, S. and Dorsey, A. T. and Fisher, Matthew P. A. and Garg, Anupam and Zwerger, W.},
  journal = {Rev. Mod. Phys.},
  volume = {59},
  issue = {1},
  pages = {1--85},
  numpages = {0},
  year = {1987},
  month = {Jan},
  publisher = {American Physical Society},
  doi = {10.1103/RevModPhys.59.1},
  url = {https://link.aps.org/doi/10.1103/RevModPhys.59.1}
}

@article{chin10,
    author = {Chin, Alex W. and Rivas, Angel and Huelga, Susana F. and Plenio, Martin B.},
    title = {Exact mapping between system-reservoir quantum models and semi-infinite discrete chains using orthogonal polynomials},
    journal = {J. Math. Phys},
    volume = {51},
    number = {9},
    pages = {092109},
    year = {2010},
    month = {09},
    issn = {0022-2488},
    doi = {10.1063/1.3490188},
}

@article{prior10,
  title = {Efficient Simulation of Strong System-Environment Interactions},
  author = {Prior, Javier and Chin, Alex W. and Huelga, Susana F. and Plenio, Martin B.},
  journal = {Phys. Rev. Lett.},
  volume = {105},
  issue = {5},
  pages = {050404},
  numpages = {4},
  year = {2010},
  month = {Jul},
  publisher = {American Physical Society},
  doi = {10.1103/PhysRevLett.105.050404},
  url = {https://link.aps.org/doi/10.1103/PhysRevLett.105.050404},
}

@article{vidal04,
  title = {Efficient Simulation of One-Dimensional Quantum Many-Body Systems},
  author = {Vidal, Guifr\'e},
  journal = {Phys. Rev. Lett.},
  volume = {93},
  issue = {4},
  pages = {040502},
  numpages = {4},
  year = {2004},
  month = {Jul},
  publisher = {American Physical Society},
  doi = {10.1103/PhysRevLett.93.040502},
  url = {https://link.aps.org/doi/10.1103/PhysRevLett.93.040502}
}

@article{yi25,
  title={A probabilistic quantum algorithm for imaginary-time evolution based on Taylor expansion},
  author={Yi, Xin and Huo, Jiacheng and Liu, Guanhua and Fan, Ling and Zhang, Ru and Cao, Cong},
  journal={EPJ Quantum Technology},
  volume={12},
  number={1},
  pages={1--22},
  year={2025},
  publisher={Springer},
  doi={doi.org/10.1140/epjqt/s40507-025-00347-0},
}

@article{sun21,
  title = {Quantum Computation of Finite-Temperature Static and Dynamical Properties of Spin Systems Using Quantum Imaginary Time Evolution},
  author = {Sun, Shi-Ning and Motta, Mario and Tazhigulov, Ruslan N. and Tan, Adrian T.K. and Chan, Garnet Kin-Lic and Minnich, Austin J.},
  journal = {PRX Quantum},
  volume = {2},
  issue = {1},
  pages = {010317},
  numpages = {14},
  year = {2021},
  month = {Feb},
  publisher = {American Physical Society},
  doi = {10.1103/PRXQuantum.2.010317},
  url = {https://link.aps.org/doi/10.1103/PRXQuantum.2.010317}
}

@Article{borrelli17,
author="Borrelli, Raffaele
and Gelin, Maxim F.",
title="Simulation of Quantum Dynamics of Excitonic Systems at Finite Temperature: an efficient method based on Thermo Field Dynamics",
journal="Scientific Reports",
year="2017",
month="Aug",
day="22",
volume="7",
number="1",
pages="9127",
issn="2045-2322",
doi="10.1038/s41598-017-08901-2",
url="https://doi.org/10.1038/s41598-017-08901-2"
}

@article{shushkov19,
    author = {Shushkov, Philip and Miller, Thomas F., {III}},
    title = {Real-time density-matrix coupled-cluster approach for closed and open systems at finite temperature},
    journal = {J. Chem. Phys.},
    volume = {151},
    number = {13},
    pages = {134107},
    year = {2019},
    month = {10},
    issn = {0021-9606},
    doi = {10.1063/1.5121749},
    url = {https://doi.org/10.1063/1.5121749},
}

@article{makri99,
  title={The linear response approximation and its lowest order corrections: An influence functional approach},
  author={Makri, Nancy},
  journal={J. Phys. Chem. B},
  volume={103},
  number={15},
  pages={2823--2829},
  year={1999},
  doi={10.1021/jp9847540},
  publisher={ACS Publications}
}

@article{allen16,
  title={Direct computation of influence functional coefficients from numerical correlation functions},
  author={Allen, Thomas C and Walters, Peter L and Makri, Nancy},
  journal={J. Chem. Theory Comput.},
  volume={12},
  number={9},
  pages={4169--4177},
  year={2016},
  doi={10.1021/acs.jctc.6b00390},
  publisher={ACS Publications}
}

@article{nusseler20,
  title = {Efficient simulation of open quantum systems coupled to a fermionic bath},
  author = {N\"u\ss{}eler, Alexander and Dhand, Ish and Huelga, Susana F. and Plenio, Martin B.},
  journal = {Phys. Rev. B},
  volume = {101},
  issue = {15},
  pages = {155134},
  numpages = {20},
  year = {2020},
  month = {Apr},
  publisher = {American Physical Society},
  doi = {10.1103/PhysRevB.101.155134},
  url = {https://link.aps.org/doi/10.1103/PhysRevB.101.155134}
}

@article{delarco25,
doi = {10.1088/2058-9565/adbdee},
url = {https://doi.org/10.1088/2058-9565/adbdee},
year = {2025},
month = {may},
publisher = {IOP Publishing},
volume = {10},
number = {3},
pages = {035018},
author = {del Arco Santos, Francisco Javier and Kottmann, Jakob S},
title = {A hybrid qubit encoding: splitting Fock space into Fermionic and Bosonic subspaces},
journal = {Quantum Sci. Technol.},
}

@article{batista01,
  title = {Generalized Jordan-Wigner Transformations},
  author = {Batista, C. D. and Ortiz, G.},
  journal = {Phys. Rev. Lett.},
  volume = {86},
  issue = {6},
  pages = {1082--1085},
  numpages = {0},
  year = {2001},
  month = {Feb},
  publisher = {American Physical Society},
  doi = {10.1103/PhysRevLett.86.1082},
  url = {https://link.aps.org/doi/10.1103/PhysRevLett.86.1082}
}

@article{fradkin89,
  title = {Jordan-Wigner transformation for quantum-spin systems in two dimensions and fractional statistics},
  author = {Fradkin, Eduardo},
  journal = {Phys. Rev. Lett.},
  volume = {63},
  issue = {3},
  pages = {322--325},
  numpages = {0},
  year = {1989},
  month = {Jul},
  publisher = {American Physical Society},
  doi = {10.1103/PhysRevLett.63.322},
  url = {https://link.aps.org/doi/10.1103/PhysRevLett.63.322}
}

@article{Whitfield11,
   title={Simulation of electronic structure Hamiltonians using quantum computers},
   volume={109},
   ISSN={1362-3028},
   url={http://dx.doi.org/10.1080/00268976.2011.552441},
   DOI={10.1080/00268976.2011.552441},
   number={5},
   journal={Molecular Physics},
   publisher={Informa UK Limited},
   author={Whitfield, James D. and Biamonte, Jacob and Aspuru-Guzik, Alán},
   year={2011},
   month=Mar, 
   pages={735–750} 
}

@article{Catto13,
doi = {10.1088/1742-6596/411/1/012009},
url = {https://doi.org/10.1088/1742-6596/411/1/012009},
year = {2013},
month = {jan},
publisher = {},
volume = {411},
number = {1},
pages = {012009},
author = {Catto, Sultan and Choun, Yoon S and Gürcan, Yasemin and Khalfan, Amish and Kurt, Levent},
title = {Grassmann Numbers and Clifford-Jordan-Wigner Representation of Supersymmetry},
journal = {J. Phys.: Conf. Ser.},
}

@article{xu24,
    author = {Xu, Xiansong and Guo, Chu and Chen, Ruofan},
    title = {Grassmann time-evolving matrix product operators: An efficient numerical approach for fermionic path integral simulations},
    journal = {J. Chem. Phys.},
    volume = {161},
    number = {15},
    pages = {151001},
    year = {2024},
    month = {10},
    issn = {0021-9606},
    doi = {10.1063/5.0226167},
    url = {https://doi.org/10.1063/5.0226167},
}

\end{document}


\title{Supporting Information: Scalable Simulation of Quantum Dynamics on Topological Quantum Hardware}
\author{Kritanjan Polley}
\email{kritanjan251@gmail.com}
\affiliation{Simons Center for Theoretical Computational Chemistry, New York University, New York, New York 10003, USA}
\author{Mark E. Tuckerman}
\email{mark.tuckerman@nyu.edu}
\affiliation{Department of Chemistry, New York University, New York, New York 10003, USA}
\affiliation{Simons Center for Theoretical Computational Chemistry, New York University, New York, New York 10003, USA}
\affiliation{Department of Physics, New York University, New York, New York 10003, USA}
\affiliation{Courant Institute of Mathematical Sciences, New York University, New York, New York 10012, USA}
\affiliation{NYU-ECNU Center for Computational Chemistry at NYU Shanghai, Shanghai 200062, China}

\maketitle

\section{Solovay-Kitaev Algorithm}\label{appSK}
Solovay-Kitaev algorithm provides an efficient method for approximating an arbitrary single qubit gate, SU(2), into a sequence of gates from a full and finite set. A brute-force approach to approximate any target unitary matrix with a given complete set of gates or basis matrices becomes exponentially expensive in the sequence length. Using the SK algorithm, the cost grows only polylogarithmically.

In order to initiate the SK algorithm, one requires an initial library of small gate sequences as a crude approximation to the target and recursively corrects its error. With Fibonacci anyons, the $B_2$ braid group generators, $\{\sigma_1, \sigma_2, \sigma_1^{-1}, \sigma_2^{-1}\}$, are shown in panel (B) of Fig.~\ref{fig2LS}. For simplicity, we restrict our discussion to $2\times2$ matrices. Now we build the initial library as
\begin{equation}
    \sigma_1, \sigma_2, \sigma_1^{-1}, \sigma_2^{-1}, \sigma_2\sigma_1^{-1}, \sigma_1\sigma_2^{-1}, \sigma_2\sigma_1^{-1}\sigma_2\cdots,
\end{equation}
up to some fixed length $l_0$. We pick the closest to our target unitary matrix ($U_{\mathrm{t}}$) from the above library, $U_0$, as zeroth order approximation or base approximation,
\begin{equation}
    U_0 \equiv SK(U_{\mathrm{t}},0)= \mathrm{BaseApproximation}(U_{\mathrm{t}}).
\end{equation}
The distance, or infidelity, between two unitary matrices is defined as
\begin{equation}
    d(U_{\mathrm{t}},U_0) = 1 - \mathrm{Tr} \left[ \big|U_{\mathrm{t}}U_0^{\dagger}\big| \right]/2 = 1 - \mathrm{Tr} \big[ |\Delta|\big]/2  \label{eqappInfidelity}.
\end{equation}

Now, we attempt to find an approximation for $U_{\mathrm{t}}$ within some error, $\mathcal{O}(\epsilon)$. At the zeroth level, where we pick the initial approximation from the initial library, the error is $d(U_{\mathrm{t}},U_0)$. In this case, we use a $k$-d thee search to find one nearest neighbor.~\cite{friedman77} If $d < \epsilon$, then the search stops here, otherwise the SK algorithm recursively corrects for the error.

The SK algorithm's fast error convergence comes from group commutator decomposition of the error matrix, $\Delta$. We can not merely replace $\Delta$ with an approximation from the initial library, as we will be replacing one $\mathcal{O}(\epsilon)$ error with a similar one. In the SK algorithm, one writes
\begin{equation}
    \Delta = VWV^{\dagger}W^{\dagger},
\end{equation}
where $V$ and $W$ are unitary matrices of distance $\mathcal{O}(\epsilon)$, but their commutator has an error $\mathcal{O}(\epsilon^2)$. We pick approximations for $V$ and $W$ from the initial library. Now, the first order approximation to the unitary looks like,
\begin{equation}
    U_1 = VWV^{\dagger}W^{\dagger} U_0.
\end{equation}
If $U_0$ has an error $\mathcal{O}(\delta)$, then it can be shown that $U_1$ has an error $\mathcal{O}(\delta^{3/2})$. This procedure can be applied recursively to obtain the desired accuracy.

\begin{figure}
    \centering
    \includegraphics[width=\linewidth]{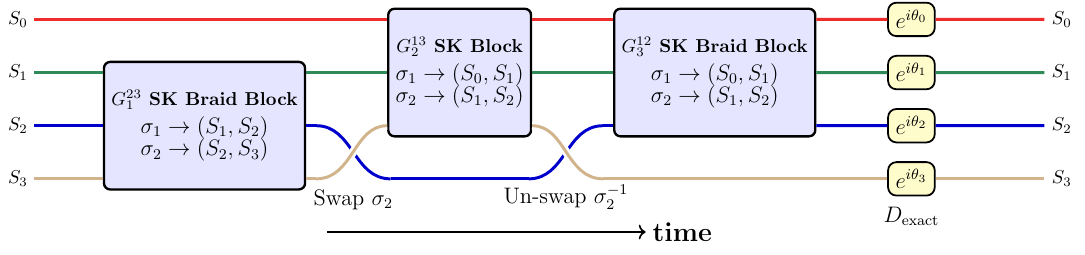}
    \caption{A schematic diagram illustrating the embedding of a propagator for a three-level Hamiltonian with anyons using Givens rotation. The decomposed gates are executed sequentially from left to right.}
    \label{fig3LSbraid}
\end{figure}

In this work, we reduced the length of the final sequence by removing adjacent inverses and using the Yang-Baxter equation,~\cite{Yang67,Baxter72} which states (\textit{c.f.} Fig.~\ref{figBraidDiagrams}),
\begin{align}
    \sigma_{j}\sigma_{k} & = \sigma_{k}\sigma_{j} ,\quad \forall \quad |j-k|\ge 2, \label{YangBaxter1}\\
    \sigma_{j}\sigma_{j+1}\sigma_{j} & = \sigma_{j+1}\sigma_{j}\sigma_{j+1}. \label{YangBaxter2}
\end{align}
Any $N$-dimensional unitary matrix can be approximated with the SK algorithm using unitary decomposition with Givens rotation and embedding. An example of unitary decomposition for a $3\times 3$ Hamiltonian is shown here, which has the form,
\begin{equation}
    H_{3\times 3} = \begin{pmatrix}
        0.0 & 0.9 & 0.1 \\
        0.9 & 0.5 & 0.0 \\
        0.1 & 0.0 & 0.8
    \end{pmatrix}.
\end{equation}
With a time step $\Delta t=0.1 \hbar$, and a target infidelity of $3.5\times 10^{-6}$ and an initial 14 word length library, the unitary decomposition has the form,
\begin{align}
    U_{\mathrm{step}} &= e^{-iH_{3\times 3}\Delta t/\hbar} = DG_3G_2G_1, \label{eqApp3LSdecom}
\end{align}
where $D$ is a diagonal matrix and $G_k$ are two-level unitaries acting on the $(2,3)$, $(1,3)$, and $(1,2)$ subspaces, respectively. Each two-level rotation has a nontrivial SU(2) block of the form
\begin{equation}
G_k\big|_{{i,j}}= \begin{pmatrix}
a_k & b_k\\
-b_k^* & a_k
\end{pmatrix}.\label{eqappGij}
\end{equation}
The elements of the three matrices in Eq.~\eqref{eqApp3LSdecom} are given in Table.~\ref{tab3LSdecomp}, and the diagonal matrix has the form, $D = \mathrm{diag}(1.0,\, e^{i0.0500677},\, e^{i0.08000133})$. The schematic of implementation of the decomposed unitaries on a quantum circuit with anyons is displayed in Fig.~\ref{fig3LSbraid}.
\begin{table}
    \centering
    \setlength{\tabcolsep}{1.8mm}
    \renewcommand{\arraystretch}{1.0}
    \label{tab3LSdecomp}
    \caption{Elements of the two-level unitary decomposition
    $U_{\mathrm{step}} = D G_3 G_2 G_1$. The form of the $G$ matrices is given in Eq.~\eqref{eqappGij}.}
    \begin{tabular}{c|c|c|c}
        \toprule
        Gate & Subspace $(i,j)$ & $a_k$ & $b_k$ \\
        \hline\hline
        $G_1$ & $(2,3)$ & $1.0$      & $0.00045 + 0.000003i$ \\
        $G_2$ & $(1,3)$ & $0.999950$ & $0.00040 + 0.009976i$ \\
        $G_3$ & $(1,2)$ & $0.995953$ & $0.00225 + 0.089844i$ \\
        \botrule
    \end{tabular}
\end{table}

\section{Cost Estimates for Fibonacci Anyons on Double Well Potential}\label{appCost}

\begin{figure}[!htb]
    \centering
    \includegraphics[width=\linewidth]{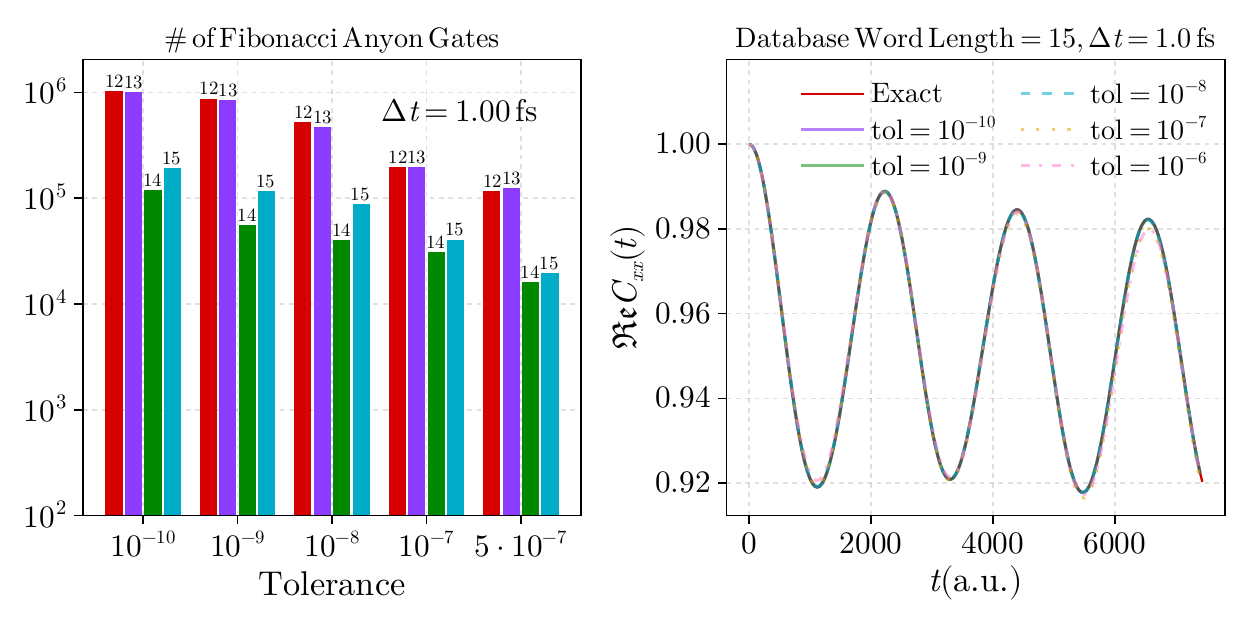}
    \caption{No. of Fibonacci gates required for simulating a double well potential, as described in Sec.~\ref{secDoubleWell}. The numbers at the top of the columns indicate the maximum word length in the initial library.}
    \label{figGateCount}
\end{figure}

We provide a simplified estimate of the number of anyon braids required for a multilevel system with double well potential, as referenced in Sec.~\ref{secDoubleWell}. The Hamiltonian size is $50\times 50$, which was decomposed into unitaries using Givens rotation. The gate count for a given target tolerance and word length of the initial library with double well potential is plotted in the left panel of Fig.~\ref{figGateCount}. For each target tolerance value, the vertical bars represent the number of anyon gates required for different word lengths of the initial library. The dynamics for an initial library word length of 15 is compared for various target tolerances on the right panel.
 
\section{Spin Boson Model with Chain-of-Bath Mapping}\label{appSB}
We want to carry out path integral calculation on this model system at finite temperature. We rewrite the bath Hamiltonian ($H_b$, Eq.~\eqref{eqSBhshb}) and the system bath coupling ($H_{sb}$, Eq.~\eqref{eqSBHsb}) for $N$ bath modes as
\begin{gather}
\mathbf{W}  = \hbar \begin{pmatrix}
\omega_1  & 0 & 0 & \cdots\\
0 & \omega_2 & 0 & \cdots\\
0 & 0 & \omega_3 & \cdots\\
\vdots & & & \ddots
\end{pmatrix}, \quad \mathbf{a} = \begin{pmatrix} 
a_1\\a_2\\\vdots\\a_N
\end{pmatrix}, \\
\mathbf{c}  = (c_1, c_2,\cdots,c_N)^{T},\\
H_b  = \mathbf{a}^{\dagger}\mathbf{W}\mathbf{a}; \quad H_{sb} = \sigma_z \Big(\mathbf{c}^{T}\mathbf{a}+\mathbf{a}^{\dagger}\mathbf{c}\Big).
\end{gather}

Now we use the properties of orthogonal polynomials for an exact unitary transformation that maps the Hamiltonian of a quantum system coupled linearly to a continuum of bosonic or fermionic modes to a Hamiltonian that describes a one-dimensional chain with only nearest-neighbor interactions.~\cite{prior10,vidal04} Following Ref.~\citenum{chin10}, we define
\begin{equation}
    \kappa = \sqrt{\mathbf{c}^{T}\mathbf{c}}, \quad\mathrm{and}\quad v_1 = \mathbf{c}/\kappa,
\end{equation}
and we arrive at the following recursive relation for unitary transformation,
\begin{align}
    \mathbf{W}v_n & = t_{n-1}v_{n-1}+\Omega_{n}v_n + t_nv_{n+1},\\
    \mathbf{V} & = (v_1, v_2, \cdots, v_N),\quad \mathbf{a} = \mathbf{Vb},\\
    H_b & = \mathbf{a^{\dagger}Wa} = \mathbf{b}^{\dagger} \big(\mathbf{V}^{\dagger}\mathbf{W}\mathbf{V}\big)\mathbf{b},
\end{align}
where
\begin{align}
W_{kk'} & = \omega_{k}\delta_{kk'}, \quad \Omega_n = v_n^{\dagger} \mathbf{W}v_n,\\
t_n & = \Vert \mathbf{W}v_n -\Omega_nv_n-t_{n-1}v_{n-1} \Vert,\\
v_{n+1} &= \frac{\mathbf{W}v_n-\Omega_nv_n-t_{n-1}v_{n-1}}{t_n}.
\end{align}

With those above relations, the spin boson Hamiltonian can be transformed as
\begin{align}
    H = & H_s + \kappa \sigma_z (b_1 + b_1^{\dagger}) \nonumber \\
    & + \hbar \sum_{n}\Omega_n b_{n}^{\dagger}b_n + \sum_{n}t_n \Big( b_{n}^{\dagger}b_{n+1} + b_{n+1}^{\dagger}b_{n}\Big),
\end{align}
where the system modes ($\sigma_z$) are coupled to the first transformed bath mode only ($b_1$ and $b_1^{\dagger}$) and the remaining bath modes are connected in a tight binding fashion. This form of spin boson model is known as ``chain-of-bath'' representation which is related to the star-bath Hamiltonian via unitary transformation.

Thermal propagation of the density matrix is inherently non-unitary in nature, which poses a challenge to the existing quantum computing algorithms. Imaginary time propagation schemes are mostly probabilistic which diminishes the probability of getting the correct measurement upon multiple applications.~\cite{sun21,yi25}

The notion of temperature in a condensed phase system enters through the bath correlation function ($C(t)$)~\cite{makri99,allen16}
\begin{align}
    C(t) = \frac{1}{\pi}\int \limits_{0}^{\infty} d\omega J(\omega) \Big[ & (n_{\beta}(\omega)+1)e^{-i\omega t}+ n_{\beta}(\omega)e^{+i\omega t} \Big],\\
    \mathrm{where} \quad n_{\beta}(\omega) & = \frac{1}{e^{\beta\hbar\omega}-1},
\end{align}
and $J(\omega)$ is the spectral density of the bath modes. We can rewrite the bath correlation function as~\cite{nusseler20}
\begin{align}
    C(t) & = \frac{1}{\pi}\int_{-\infty}^{\infty} d\omega J_{\beta}(\omega)e^{-i\omega t},\\ \nonumber \\
    J_{\beta}(\omega) & = \begin{cases}
    \frac{J(\omega)}{1-e^{-\beta \hbar \omega}}, & \omega>0\\
    \frac{J(|\omega|)}{e^{\beta \hbar |\omega|} -1}, & \omega< 0
    \end{cases},
\end{align}

With the above transformation, we transform all temperature dependence on the spectral density function itself. We sample the coupling constants and frequencies from the temperature dependent spectral density, the coupling constants will carry the notion of temperature and the rest of the system will act like a zero Kelvin system. The exact but alternative finite temperature formulation of quantum mechanics is known as thermofield dynamics.~\cite{borrelli17,shushkov19}

We split the Hamiltonian into smaller blocks which is intuitive from the chain-bath picture. We trotterized the Hamiltonian in first order in the above equations, a second order symmetrized Suzuki-Trotter is used in the calculations.
\begin{align}
e^{-iH\Delta t} \approx & \prod_{j\in \mathrm{odd}} e^{-ih_{j,j+1}\Delta t/\hbar} \prod_{j\in \mathrm{even}} e^{-ih_{j,j+1}\Delta t/\hbar}, \label{eqOddEvenSplitFull}\\
h_{0,1} & = H_s + \hbar \Omega_1 b_1^{\dagger} b_1 + \kappa \sigma_z ( b_1^{\dagger} + b_1), \label{eqOddEvenSplit}\\
h_{j,j+1} & = \hbar\Omega_j b_j^{\dagger} b_j + t_j \big(b_{j}^{\dagger} b_{j+1} + b_{j+1}^{\dagger} b_{j}\big).\label{eqOESform}
\end{align}

If we choose $d$ Fock levels for each bath mode, then the dimension for $h_{0,1}$ is $2d\times 2d$ and the rest $h_{j,j+1}$ have a size of $d^2\times d^2$. Now, we define the total wavefunction with $M+1$ indices, where we have 1 system mode and $M$ coupled bath modes.
\begin{align}
    |\alpha_{n}\rangle = & |\sigma_n, m_{1,n}, \dots,m_{M,n}\rangle,
\end{align}
where $n$ is the time slice index. We take expectation value of the propagator for time $t$ by inserting $N$ identities in the trotted split form such that $\Delta t=t/N$.
\begin{align}
    \langle \alpha_N | e^{-iHt/\hbar}| \alpha_0 \rangle \approx & \sum_{\alpha_1, \dots, \alpha_{N-1}} \prod_{n=0}^{N-1} \langle \alpha_{n+1} | e^{-iH \Delta t/\hbar}| \alpha_{n} \rangle. \label{eqFullPropagator}
\end{align}

Next, we insert the trotterized form of $e^{-iH\Delta t}$ from Eqs.~\eqref{eqOddEvenSplit}-\eqref{eqOESform} into Eq.~\eqref{eqFullPropagator} and introduce further identities between blocks of the chain-of-bath Hamiltonian,  
\begin{align}
\langle \alpha_{n+1} | e^{-iH \Delta t/\hbar}| \alpha_{n} \rangle  \approx & \sum_{\overline{\alpha}_n} \prod_{j\in \mathrm{odd}} \langle \overline{\alpha}_{j,n} | U_{j,j+1}|\alpha_{j,n} \rangle \nonumber\\
& \prod_{j\in \mathrm{even}} \langle \alpha_{j,n+1} | U_{j,j+1}| \overline{\alpha}_{j,n} \rangle,
\end{align}
where $U_{j,j+1}\equiv\exp \{ -ih_{j,j+1}\Delta t /\hbar\}$. We pick the initial state where the system and bath modes are uncorrelated. In the temperature dependent spectral density setup, the bath modes are sitting in their lowest states, while the coupling constants are carrying the notion of temperature.
\begin{equation}
    |\psi(0)\rangle = |1\rangle_s \otimes |0\rangle_{b_1} \otimes \cdots \otimes|0\rangle_{b_M},
\end{equation}
with coefficients
\begin{equation}
    \psi_{s,n_1,n_2\cdots n_M}(0) = \delta_{s,1}\prod_{j=1}^{M} \delta_{n_j,0}. 
\end{equation}

The total propagator is split as
\begin{equation}
    U(\Delta t) \approx U_{\mathrm{odd}}(\Delta t/2) U_{\mathrm{even}}(\Delta t) U_{\mathrm{odd}} (\Delta t/2),
\end{equation}
and the total propagation has the form
\begin{align}
    \psi_{\alpha_{n+1}} &  =   \sum_{\alpha_n, \beta_n, \gamma_n} \Big\langle \alpha_{n+1} \Big| U_{\mathrm{odd}}(\Delta t/2) \Big| \gamma_{n} \Big\rangle \Big \langle \gamma_{n} \Big| U_{\mathrm{even}}(\Delta t) \Big| \beta_{n} \Big\rangle \Big\langle \beta_{n} \Big| U_{\mathrm{odd}}(\Delta t/2) \Big| \alpha_{n} \Big\rangle \psi_{\alpha_{n}}.
\end{align}
The tensor network algorithm does not evaluate the path integral trajectories explicitly. Instead, each application of a two-site gate contracts a local tensor into the matrix product state, and the matrix product bond indices compactly encode the coherent sum over exponentially many partial paths. This tensor network formalism efficiently compresses the full path integral. These small unitary matrix multiplications are carried out with topological braids.

\section{Fermionic Path Integral on Second Quantized Hamiltonians}\label{appGrassmann}
Dynamics of fermionic particles can be described using Grassmann variables,~\cite{Altland_Simons_book} also known as anticommuting or fermionic variables. A $N$-dimensional Grassmann algebra $G\equiv\{\theta_j\}$ is formed by generators $\theta_j$ with $j=1,\dots,N$ that satisfy
\begin{equation}
    \theta_{j}\theta_{k} + \theta_{k}\theta_{j}=0, \quad \mathrm{and} \quad \theta_j^2 = 0.
\end{equation}

Integrals of Grassmann variables, known as Berezin integrals,~\cite{peskinBook} are defined to be identical to differentiation,
\begin{align}
    \int d\theta \equiv \frac{\partial}{\partial \theta}.
\end{align}
A few examples of integration with Grassmann variables are given below,
\begin{gather}
    \int d\theta  = 0,\quad \int d\theta \, \theta  =1,\, \int d\theta_1 d\theta_2 e^{-a\theta_1\theta_2}  = a.
\end{gather}
The last relation can be extended for $N$ variables as
\begin{equation}
    \int \mathcal{D}\overline{\theta} \mathcal{D}\theta \exp \bigg \{ -\sum_{j,k=1}^N \overline{\theta}_j A_{jk}\theta_k \bigg\} = \mathrm{det} (A),
\end{equation}
where $\mathcal{D}\theta\equiv d\theta_1\dots d\theta_N$.

Our goal is to compute the time correlation function, which, for simplicity, we carry out at zero temperature,
\begin{align}
    C_{AB}(t) & =  \langle 0 | e^{iHt} A e^{-iHt} B | 0 \rangle, \label{eqCabt}
\end{align}
where $H$ is Hamiltonian governing the system and $A$ and $B$ are two system operators. Now, we consider $N$ fermionic modes with creation and annihilation operators,
\begin{equation}
    \{c_j, c_k^{\dagger}\} =\delta_{jk}, \quad \{c_j,c_k\}=\{c_j^{\dagger}, c_k^{\dagger}\}=0.
\end{equation}
We introduce two sets of independent Grassmann variables, $\theta_j$ and $\overline{\theta}_j$. The fermionic coherent ket and bra states are defined by
\begin{align}
    |\theta\rangle & = e^{-\sum_j \theta_jc_j^{\dagger}}|0\rangle,\quad  \langle\overline{\theta}| = \langle 0|e^{-\sum_j c_j\overline{\theta}_j},
\end{align}

The identity involving the coherent states is given by,
\begin{equation}
    1 = \int \prod_{j=1}^{N}d\overline{\theta}_j d\theta_j e^{-\overline{\theta}\theta} |\theta \rangle \langle \overline{\theta} |. \label{eqIdentity}
\end{equation}
Now we introduce trotter splitting for the propagator in Eq.~\eqref{eqCabt},
\begin{equation}
    e^{-iHt} = \Big(e^{-iH\Delta t} \Big)^P, \quad \Delta t = t/P,
\end{equation}
\begin{align}
    & C_{AB}(t)  = \int \bigg(\prod_k d\overline{\theta}_k d\theta_k\bigg) e^{-\sum_k \overline{\theta}_k \theta_k} \langle 0 | \theta_{f} \rangle \langle \overline{\theta}_{f} | e^{iH\Delta t} \cdots A \cdots e^{-iH\Delta t} B | \theta_{i} \rangle \langle \overline{\theta}_{i} | 0\rangle,
\end{align}
where $\cdots$ indicate the remaining $P-1$ copies of the short time propagator with inserted identity. For the vacuum Fock state, $\langle \overline{\theta}_{i} | 0 \rangle = \langle 0 | \theta_{f} \rangle = 1$. For each individual short time propagators, we get
\begin{align}
    & e^{-\overline{\theta}_k \theta_{k}} e\langle \overline{\theta}_k | e^{-iH\Delta t}| \theta_{k-1} \rangle  \nonumber \\
    & \approx e^{-\overline{\theta}_k \theta_{k}}\exp\big\{\overline{\theta}_k \theta_{k-1} - i\Delta t H(\overline{\theta}_k, \theta_{k-1})\big\}, \\
    & = \exp\big\{\overline{\theta}_k (\theta_{k-1} -\theta_k)- i\Delta t H(\overline{\theta}_k, \theta_{k-1})\big\}.
\end{align}

Summing over all intervals produces,
\begin{equation}
    \exp \left\{ \sum_{k=1}^P \Big[\overline{\theta}_k (\theta_{k-1}-\theta_k) -i\Delta t H(\overline{\theta}_k, \theta_{k-1})\Big] \right\}.
\end{equation}

Putting the propagator factors and operator insertions together gives
\begin{align}
& C_{AB}(t) = \lim_{P\rightarrow\infty} \int \prod_{k=1}^{2P+1} d\overline{\theta}_k\,d\theta_k\; A(\overline{\theta}_{P+1},\theta_P) B(\overline{\theta}_1,\theta_0) \nonumber\\
&\times \exp\Bigg\{ \sum_{k=1}^{P} \Big[ \overline{\theta}_k(\theta_{k-1}-\theta_k) -i\Delta t\, H(\overline{\theta}_k,\theta_{k-1})
\Big] \nonumber\\
&+ \sum_{k=P+2}^{2P+1} \Big[ \overline{\theta}_k(\theta_{k-1}-\theta_k) +i\Delta t\, H(\overline{\theta}_k,\theta_{k-1}) \Big] \Bigg\}, \\
& =\int \mathcal{D} \overline{\theta} \mathcal{D} \theta \,\, A\big(\overline{\theta}(t), \theta(t)\big) B\big(\overline{\theta}(0), \theta(0)\big) e^{iS[\overline{\theta}, \theta]},
\end{align}
where 
\begin{align}
    A(\overline{\theta}_{P+1},\theta_P) & = e^{-\overline{\theta}_{P+1}\theta_{P}} \langle \overline{\theta}_{P+1}| A | \theta_{P} \rangle.
\end{align}

The one-body contribution to the electronic Hamiltonian is
\begin{equation}
    H_1 = \sum_{j,k} h_{jk} (\bm{R}) c_j^\dagger c_k,
\end{equation}
where $h_{jk} (\bm{R})$ is the single particle Hamiltonian matrix. Instead of manipulating anti-commuting creation ($c_j^\dagger$) and annihilation ($c_k$) operators, we map them directly onto a conjugate pair of Grassmann fields, $\overline{\theta}_j$ and $\theta_k$. The quadratic single-particle Hamiltonian operator $H_1$ translates into a pure Grassmann function:
\begin{equation}
H_1(\overline{\theta}, \theta) = \sum_{j,k} h_{jk}(\bm{R}) \overline{\theta}_j \theta_k
\end{equation}

We define the multi-orbital fermionic coherent states using the boundary fields $\langle \overline{\theta}_f |$ for the final state and $| \theta_i \rangle$ for the initial state. $N$-electron Slater determinants $|I\rangle$ and $|J\rangle$ project into this Grassmann basis as ordered products of variables,
\begin{align}
\langle \overline{\theta}_f | I \rangle & = \overline{\theta}_{f, i_1} \cdots \overline{\theta}_{f, i_N}, \,\, \langle J  | \theta_i \rangle = \theta_{i, j_N} \cdots \theta_{i, j_1}.
\end{align}

Because the action generated by $H_1(\overline{\theta}, \theta)$ is strictly quadratic, evaluating the matrix element of the evolution operator $U_1(\Delta t) = e^{-iH_1\Delta t}$ between the initial and final coherent states yields an exact Gaussian exponential weight,
\begin{equation}
\langle \overline{\theta}_f | e^{-iH_1 \Delta t} | \theta_i \rangle = \exp \left( \sum_{j,k} \overline{\theta}_{f, j} [U_1(\Delta t)]_{jk} \theta_{i, k} \right),
\end{equation}
where $U_1(\Delta t)$ is the standard single-particle matrix exponential. To compute the full many-body transition amplitude $\langle I | e^{-iH_1 \Delta t} | J \rangle$, we insert the resolution of identity,
\begin{align}
\langle I | e^{-iH_1 \Delta t} | J \rangle &= \int \mathrm{d}\overline{\theta}_f\, \mathrm{d}\theta_f\, \mathrm{d}\overline{\theta}_i\, \mathrm{d}\theta_i\, e^{-\overline{\theta}_f\theta_f} e^{-\overline{\theta}_i\theta_i} \langle I | \theta_f \rangle \exp \left\{ \overline{\theta}_f^{T} U_1(\Delta t) \theta_i \right\} \langle \overline{\theta}_i | J \rangle \\
& = \mathrm{det} \big(\left[ U_{1}(\Delta t)\right]_{i_a,j_b}\big)_{a,b=1}^{N}.
\end{align}

The density-density Coulomb repulsion operator mapped to Grassmann variables,
\begin{equation}
H_2 = \sum_{j>k} U_{jk} (\bm{R}) n_j n_k, \quad n_j = c_j^{\dagger}c_j.
\end{equation}
To get the Hubbard-Stratonovich (HS) transformation,~\cite{Altland_Simons_book} we use
\begin{equation}
    e^{\alpha n_j n_k} = \frac{1}{2} \sum_{s_{jk}=\pm 1} e^{(c_{jk}+s_{jk}\lambda_{jk})(n_j+n_k)}.
\end{equation}
For a given value of $\alpha$, we can solve for two unknowns ($c_{jk}$ and $\lambda_{jk}$) from two equations. They will have form
\begin{align}
    \lambda_{jk} & = \mathrm{arccosh}(1.0 / \sqrt{2.0 - \exp{(\alpha)}}), \nonumber \\
    c_{jk} & = -\log{(\cosh{\lambda_{jk}})},
\end{align}
and once we write them down as effective single body terms, we use the same procedure as used for one-body terms. Note, $\lambda_{jk}$ and $c_{jk}$ are, in general, complex. 

$H_1$ is quadratic, while after the Hubbard–Stratonovich transformation each fixed auxiliary-field configuration of $H_2$ is represented by an effective quadratic one-body propagator. Therefore $H_1$ and the HS transformed $H_2$, for fixed auxiliary field, have Gaussian coherent-state kernels. Evaluated between arbitrary coherent state boundaries, they yield exact Gaussian matrix kernels,
\begin{gather}
\langle \overline{\theta}_A | e^{-i H_1 \Delta t} | \theta_B \rangle = \exp\left( \overline{\theta}_A^T e^{-i h^{(1)} \Delta t} \theta_B \right),\\
\langle \overline{\theta}_A | e^{-i H_2 \Delta t} | \theta_B \rangle = \frac{1}{2}\sum_{s_k}\exp\left( \overline{\theta}_A^T U_2(s_k) \theta_B \right),\\
\quad [U_2(s_k)]_{pq} = \delta_{pq} \exp \Big\{\sum_{r\neq p} (s_{pr}\lambda_{pr} + c_{pr}) \Big\},\\
\langle \overline{\theta}_A | e^{-i H_3 \Delta t} | \theta_B \rangle =
\sum_{I, J} \Big(\overline{\theta}_{A, i_1} \dots \overline{\theta}_{A, i_N}\Big) \big[ U_3(\Delta t) \big]_{IJ} \Big(\theta_{B, j_N} \dots \theta_{B, j_1}\Big).
\end{gather}
$U_3(\Delta t) = e^{-i H_3 \Delta t}$, the exchange part of the Hamiltonian, is dense and non-diagonal due to the non-local exchange integrals, requiring explicit numerical matrix exponentiation.

To link these three sequential steps within the $k$-th time slice, we insert two identities in the Grassmann coherent state basis, tracking the intermediate cross-states via fields labeled 1 and 2,
\begin{align}
\mathcal{I}_1= & \langle \overline{\theta}_k | e^{-i H \Delta t} | \theta_{k-1} \rangle \approx \int \mathcal{D}(\overline{\theta}_{k,2}, \theta_{k,2}) \mathcal{D}(\overline{\theta}_{k,1}, \theta_{k,1}) \nonumber \\
& \times e^{-\overline{\theta}_{k,2}^T \theta_{k,2} - \overline{\theta}_{k,1}^T \theta_{k,1}}  \langle \overline{\theta}_k | e^{-i H_3 \Delta t} | \theta_{k,2} \rangle \langle \overline{\theta}_{k,2} | e^{-i H_2 \Delta t} | \theta_{k,1} \rangle \langle \overline{\theta}_{k,1} | e^{-i H_1 \Delta t} | \theta_{k-1} \rangle.
\end{align}

The Grassmann integral over the first intermediate cross-state pair ($\overline{\theta}_{k,1}, \theta_{k,1}$) is evaluated as,
\begin{align}
& \langle \overline{\theta}_{k} | e^{-i H_2 \Delta t} e^{-i H_1 \Delta t} | \theta_{k-1} \rangle |_{s_k} \nonumber\\
& = \int \mathcal{D}\overline{\eta} \mathcal{D}\eta e^{-\overline{\eta}^{T}\eta} \langle \overline{\theta}_k | e^{-i H_2 \Delta t} | \eta \rangle \langle \overline{\eta} | e^{-i H_1 \Delta t} | \theta_{k-1} \rangle \nonumber \\
&= \exp \left( \overline{\theta}_{k}^T U_2(s_k) U_1(\Delta t) \theta_{k-1} \right).
\end{align}
Substituting the explicit Gaussian kernels transforms the integrand into a coupled field expression,
\begin{align}
& \big\langle I_k \big| e^{-i H \Delta t} \big| I_{k-1} \big\rangle  = \frac{1}{2}\sum_{K_k,s_k} \big[ U_3(\Delta t) \big]_{I_k K_k} \det \Big( \big[ U_2(s_k) U_1(\Delta t) \big]_{(K_k)_a, ({I_{k-1}})_b} \Big)_{a,b=1}^{N}.
\end{align}                          

We define,
\begin{align}
\big[\mathcal{U}_k\big]_{I_k I_{k-1}} & = \frac{1}{2}\sum_{\mathbf{s}_k} \sum_{K_k}  \big[U_3\big]_{I_k K_k} \det\left( \big[U_2(\mathbf{s}_k)U_1\big]_{(K_k)_a,(I_{k-1})_b} \right)_{a,b=1}^{N},
\end{align}
and, with that definition, the final propagator has the form,
\begin{equation}
\big\langle I \big| \hat{U}_{\mathrm{total}} \big| J \big\rangle = \sum_{I_1,\ldots,I_{P-1}} \prod_{k=1}^{P} \big[\mathcal{U}_k\big]_{I_k I_{k-1}}, \, I_0=J, I_P=I.
\end{equation}

\bibliography{reference}